\documentclass[a4paper,11pt]{article}
\usepackage{jinstpub} 
\usepackage{lineno}

\usepackage{subcaption}
\usepackage{dirtytalk}
\usepackage{array}
\usepackage{booktabs}
\usepackage[export]{adjustbox}

\title{\boldmath Construction and Performance of the sMDT Precision Muon Tracking Chambers for ATLAS at the HL-LHC}

\author[]{D. A. Buchin,}
\author[]{O. Kortner,}
\author[1]{H. Kroha,}\note{Corresponding author.}
\author[]{A. Reed,}
\author[]{M. Rendel,}
\author[]{P. Rieck,}
\author[]{E. Voevodina,}
\author[]{and V. M. Walbrecht}
\affiliation[]{Max Planck Institute of Physics, Boltzmannstraße 8, Garching, Germany}

\emailAdd{kroha@mpp.mpg.de}

\abstract{In order to improve the muon trigger efficiency and the rate capability of the ATLAS muon detectors for operation at the high luminosity upgrade of the Large Hadron Collider (HL-LHC), the Monitored Drift Tube (MDT) tracking chambers in the inner barrel layer of the ATLAS Muon Spectrometer will be replaced by small-diameter Muon Drift Tube (sMDT) chambers integrated with new thin-gap RPC trigger chambers. The sMDT chambers are in serial production since January 2021. This documentation provides insight into the drift tube production and every step performed in the chamber construction at the MPI Munich. The serial production involves a stringent quality control program to assure the reliability and high mechanical precision of the chambers. This program consists of tests of the individual drift tubes, measurements of the chamber geometry and proof of gas tightness. Final certification at the production site is provided using cosmic rays. The dedicated quality control database and monitoring Web interface that is used commonly for both production sites of the sMDT chambers is presented.}


\begin{document}
\maketitle
\flushbottom

\section{Introduction}
\label{sec:intro}
The barrel part of the muon spectrometer of the ATLAS experiment during the third run of the LHC~\cite{PERF-2007-01,GENR-2019-02} consists of Monitored Drift Tube (MDT) and small-diameter Muon Drift Tube (sMDT) chambers~\cite{sMDT_concept,sMDT_design} for precision muon tracking and of Resistive Plate Chambers (RPC) for muon triggering. The precision tracking chambers are arranged in three radial layers an 16 azimuthal sectors around the beam axis with alternating large and small sectors. Two layers of RPCs are situated in the middle and one in the outer spectrometer layer in combination with the MDT chambers.

\begin{figure}[htb]
    \centering
    \includegraphics[width=0.95\textwidth]{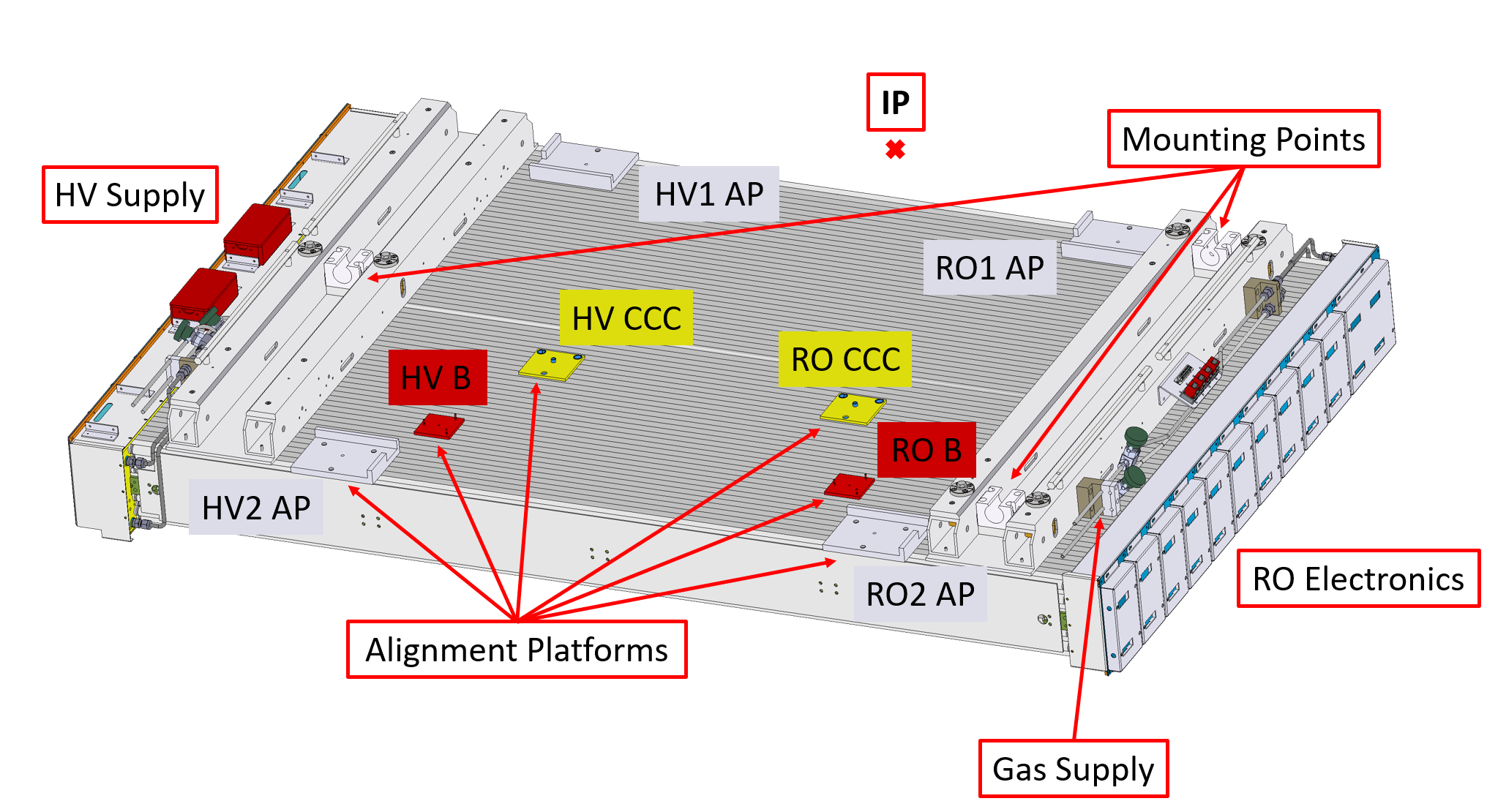}
    \caption{3D drawing of an A-side BIS1 sMDT chamber. The chamber ends are equipped with the High-Voltage (HV) distribution and the readout (RO) electronics The orientation  with respect to the interaction point (IP) inn ATLAS is indicated with the drift tubes perpendicular to the beam axis. The supports with rail bearings (two-point on the RO side and single-point on the HV side) are glued to upper tube layer during assembly oriented radially outward to the toroid magnet coils of the muon spectrometer to which they are mounted. This outer tube layer also carries mounting platforms for optical alignment sensors ov various types (see text.}
    \label{fig:sMDT_chamber_CAD}
\end{figure}

The ATLAS detector will be upgraded for operation at the High-Luminosity LHC (HL-LHC) in the Long Shutdown 3 (LS3) of the LHC between 2026 and 2030. The ATLAS muon spectrometer will undergo major upgrades in order to cope with the increased data rates~\cite{ATLAS-TDR-26}. Additional thin-gap RPCs will be added in the inner barrel layer of the spectrometer, integrated with the sMDT chambers in common modules, to improve the trigger acceptance and efficiency. The MDT chambers in the innermost barrel layer in the small sectors, named BIS chambers, will be replaced by sMDT chambers (see Figure~\ref{fig:sMDT_chamber_CAD}) in order to accommodate the additional RPCs in the available space and to increase the background rate capability in this region of highest neutron and $\gamma$ ray background radiation in the barrel muon detector, especially at the HL-LHC with roughly ten times higher expected background rates in the muon detector compared to the LHC. 

In addition, both the legacy MDT chambers and the sMDT chambers will be equipped with new front-end electronics, including new Amplifier-Shaper-Discriminator (ASD)~\cite{ASD2_IEEESensor,ABOVYAN2019374,ASD2_manual} and TDC ASICs~\cite{GUO2021164896} which support the higher data rates and continuous readout for the (s)MDT-based first-level muon trigger for HL-LHC~\cite{ATLAS-TDR-26}. The sMDT chambers consist of drift tubes of 15~mm diameter, which is half the diameter of the drift tubes of the legacy MDTs. The maximum drift time of the sMDT tubes is only 175~ns compared to about 720~ns for the legacy MDT tubes leading, together with the smaller cross section exposed to background radiation, to about eight times lower occupancy. 

The effect of background induced space charge fluctuations in the drift tubes on the spatial resolution sets in only at drift distances larger than the sMDT tube radius. Gas gain loss caused by space charge shielding the electrical field strength in the avalanche region near the sense wire is suppressed proportional to the third power of the drift tube radius. Altogether, the rate capability is improved by about an order of magnitude~\cite{First_sMDT_prototype_for_NSW,
sMDT_design_performance_conf2015,
sMDT_design_performance_conf2016,
sMDT_electronics_performance}. Because of signal pile-up effects deteriorating drift tube efficiency and spatial resolution at high counting rates, the rate capability also depends on the readout electronics. These effects have been studied for the new ASD readout chip which improves the rate capability due to its 3~ns faster peaking time a lower noise rate and faster baseline return~\cite{sMDT_electronics_performance}. 

Prototypes of sMDT chambers for the muon detector barrel region (rectangular shape) and for the endcap regions (trapezoidal shape) have been constructed and tested in muon beams and under background radiation since 2010~\cite{First_sMDT_prototype_for_NSW}. sMDT and MDT chambers share the requirement of high mechanical precision with sense wire positioning accuracy of better than $20~\mu$m. The sMDT chamber concept and design described below improved the precision and simplified the assembly at lower cost and construction time for large-scale production.

Several types of sMDT chambers have been designed and built over the past years and installed in the ATLAS detector in 2013 and 2016 to increase the acceptance and the spatial resolution of the ATLAS muon spectrometer~\cite{sMDT_BME_BMG_ATLAS,sMDT_BMG,sMDT_BMG_ATLAS,
sMDT_BMG_BIS_ATLAS,sMDT_design_performance_conf2016}. As a pilot project for the Phase 2 upgrade of ATLAS in LS3, 16 sMDT chambers have been produced to replace the MDTs at the edges of the small inner barrel sectors, the so called BIS78 chambers~\cite{sMDT_BIS78,sMDT_RPC_BIS78,sMDT_BIS_ATLAS}. The 8 BIS78 chambers for one hemisphere of the detector (A-side) were installed with the integrated thin-gap RPCs in the Long Shutdown 2 (LS2) of the Phase 1 upgrade of the LHC in 2020--2021 before the start of the latest LHC Run~3~\cite{sMDT_BIS78_commissioning_installation,GENR-2019-02}. The remaining BIS78 chambers for the C-side hemisphere will only be installed with their thin-gap RPCs in LS3. 

\begin{figure}[htb]
    \centering
    \includegraphics[width=0.85\textwidth]{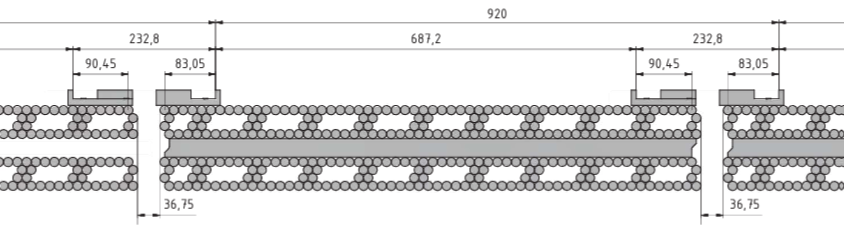}
    \caption{Layout of a representative section of a BIS1-6 sMDT chamber layer aligned showing the dimensions and close distances between neighbouring chambers.} 
    \label{fig:BIS_layer}
\end{figure}

The new BIS1-6 sMDT chambers to replace the remaining MDTs in the small sectors of the inner barrel for HL-LHC have been designed at Max Planck Institute for Physics in Munich (MPI Munich) where the sMDT chambers have been developed including the readout electronics (see Figures~\ref{fig:sMDT_chamber_CAD} and \ref{fig:BIS_layer}). The production of the 96 BIS1-6 chambers for the Phase 2 upgrade was equally shared between two construction sites, one at the MPI Munich for A-side chambers and one at the University of Michigan for C-side chambers and completed between January 2020 and September 2023.
The BIS1 chambers consist of 560 and the BIS2-6 chambers of 464 drift tubes in two densely packed multilayers of four tube layers each separated by an aluminum spacer frame of only 30.6~mm height which carries an optical planarity monitoring system. Mounting platforms for the optical sensors of the global ATLAS muon spectrometer alignment system are glued with high precision on the tube layers. The geometrical parameters and properties of the BIS1-6 sMDT chambers are summarized in Tables~\ref{tab:BIS1-6_parameters} and \ref{tab:BIS1-6_summary}. 

In this paper, the construction as well as the quality control and performance tests of the 48 BIS1-6 sMDT chambers for the A-side hemisphere of ATLAS at the MPI Munich between January 2020 and December 2022. Four additional spare chambers of that type have been constructed as well. 

\begin{table}
\centering

\caption{Geometry parameters of the BIS~1-6 sMDT 
chambers~\cite{ATLAS-TDR-26,sMDT_ATLAS_parameter_book}}
\label{tab:BIS1-6_parameters}
\begin{tabular}{|l|c|c|}
\hline
Type & BIS~1 & BIS~2-6 \\
\hline
Number of chambers & 16 & 80 \\
Chamber width in beam dir. (mm)  & 1097 &  916 \\
Chamber length (mm)        & 1796 & 1796 \\
Aluminum tube length (mm)  & 1615 & 1615 \\
Assembled tube length (mm) & 1624 & 1624 \\
Active area/chamber (m$^2$)& 1.64 & 1.36 \\
\hline
Tube layers & $2\times 4$ & $2\times 4$ \\
Tubes/layer & 70 & 58 \\
Tubes/chamber & 560 & 464 \\
Spacer height (mm)  & 30.6 & 30.6 \\
Multilayer height (mm)   & 139 & 139 \\
Chamber height (mm) & 249 & 249 \\
\hline
Gas volume/chamber (liter) & 139.4 & 115.5 \\
Chamber weight (kg)    & 140 &  115 \\
\hline
CSMs/chamber & 2 & 1 \\
Mezzanine cards (24 ch.)/chamber  & 24 & 20 \\
Mezzanine cards/CSM1 & 12 & 20 \\
Mezzanine cards/CSM2 & 12 & 0  \\
Temperature sensors/chamber             & 10 & 10 \\
B-field sensors/chamber       &  2 &  2 \\
Praxial alignment platforms/chamber &  4 &  4 \\
CCC alignment platforms/chamber &  2 &  1-2 \\
\hline
\end{tabular}
\end{table}
\begin{table}
\centering
\caption{BIS~1-6 sMDT chambers 
summary~\cite{sMDT_ATLAS_parameter_book}}
\label{tab:BIS1-6_summary}
\centering
\begin{tabular}{|l|c|}
\hline
Parameter & Value \\
\hline
Number of chambers & 96 \\
Number of tubes & 46080 \\
Total tube and wire length & 74.5~km \\ 
Total chamber active area & 135~m$^2$ \\
Total gas volume & 114.7~m$^3$ \\
Total chamber weight & 11.5 tons \\
Number of mezzanine cards & 1984 \\
Number of hedgehog boards & 3968 \\
Number of CSMs & 192 \\
\hline 
\end{tabular}
\end{table}

\section{Chamber Construction}
\label{sec:Chambers}
The construction steps of the BIS1-6 sMDT chambers, equal for both production sites, are described in this section. 
Detailed descriptions of the construction campaign of A-side chambers at MPI Munich are provided in~\cite{Kroha:2024lod,Rendel:2023,Buchin:2026}. The production of drift tubes started in 2019, the first BIS1 chamber was completed in January 2020.
The construction of the BIS1-6 C-side chambers is reported in~\cite{ChamberConstruction_UM_2023} and the corresponding drift tube production and test in~\cite{TubeProduction_UM_2022}.

\subsection{Drift Tube Design and Production}
\label{sec:Chambers:Tubes}
\begin{figure}[htb]
    \centering
      \includegraphics[width=0.95\textwidth]{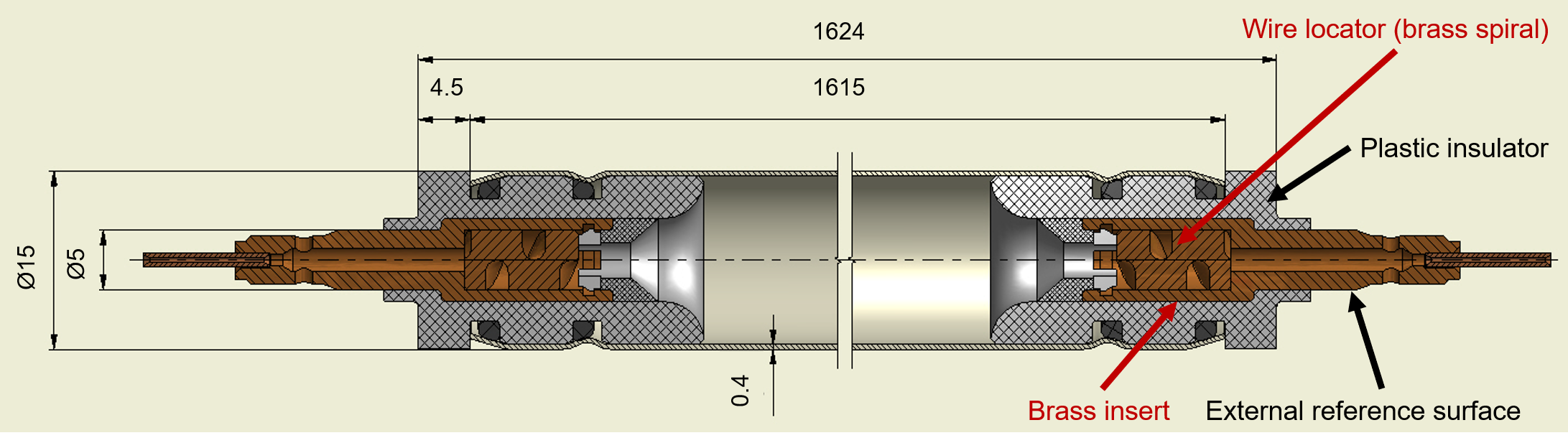}
    \caption{Drift tube design and dimensions. The wire positioning concept by means of a spiral wire locator and the cylindrical external reference surface, both on the same brass insert of the endplug, is illustrated. All dimensions are in mm.} 
    \label{fig:TubeDesign}
\end{figure}

The design and dimensions of the BIS1-6 sMDT drift tubes and their endplugs is illustrated in Figure~\ref{fig:TubeDesign}.
The drift tube design parameters, equal for all BIS1-6 sMDT chambers, are summarised in Table~\ref{tab:Tube_Design_Parameters}. 
The raw aluminum tubes have been extruded with high precision by MIFA Aluminium in the Netherlands. Their straightness deviation is $\leq 0.5$~mm over a tube length of 1600~mm as verified on a random sample of delivered tubes. The raw tubes were chemically etched and chromatised using SurTec 650 tri-chrome passivation on the inside and outside for cleaning and to ensure reliable electrical ground contacts. 

The 50~$\mu$m diameter gold-plated tungsten-rhenium (97:3) sense wires are fixed at either end of the drift tubes in an endplug. An endplug consists of a injection molded polybutylene terephthalate (PBTP) plastic insulator with 30\% glass fiber reinforcement as main body with a brass insert holding in a fitting bore a brass spiral as wire locator which is held in place by a PBTP stopper. PBTP was selected as insulator material, and also for the gas connectors to the tubes, because of its low outgassing property and insensitivity to cracking, both even under very high irradiation. The aging properties of the sMDT drift tubes have been tested under long-term high-intensity $\gamma$ irradiation at the nominal gas gain of $2\cdot 10^{4}$, including the endplugs, showing no sign of aging up to 14~C/cm charge accumulation on the sense wires~\cite{Hertenberger:2024jcj}, a factor of 23 more than required at the LHC and a factor of 2.5 above the requirements at HL-LHC.

\begin{figure}[htb]
    \centering
    \includegraphics[width=0.95\textwidth]{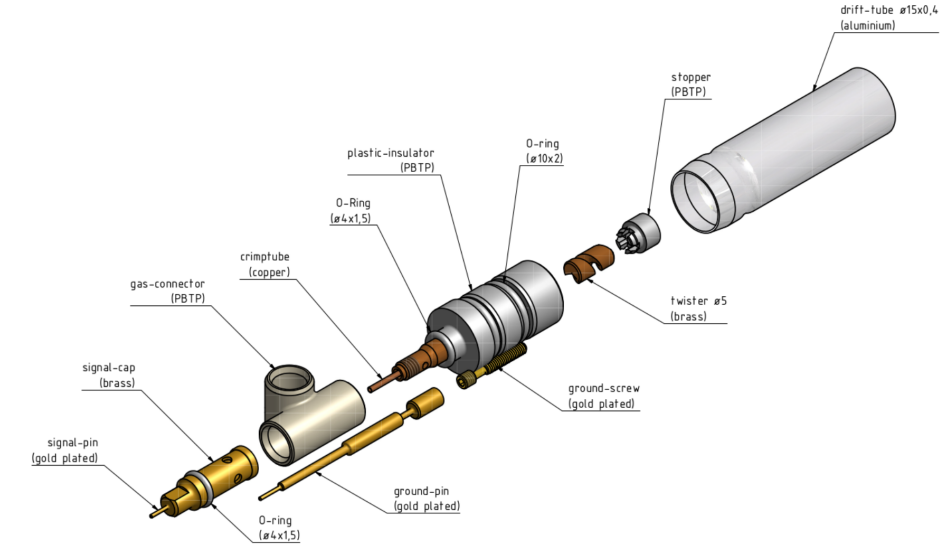}
    \caption{Exploded view of the sMDT endplugs with their components including the connectors to the gas distributions system (see Section~\ref{sec:Chamber:Gas}) and the grounding screws inserted between adjacent tubes. All dimensions are in mm.}
    \label{fig:EndplugDesign}
\end{figure}

The spiral wire locator with a central bore hole of $50~\mu$m fitting the sense wire positions it with respect to the cylindrical reference surface on the brass insert on the outside of the endplug with a precision of order $1\mu$m corresponding to the machining precision of the brass insert (see Figure~\ref{fig:TubeDesign}). The spiral design allows for the automated feeding of the wire through the endplug. This wire locator concept was already used for the endplugs of the MDT drift tubes. The innovation for the sMDT endplugs is the external reference surface directly on the brass insert which is inserted into bore holes the combs of the chamber assembly jigging, increasing the sense wire positioning accuracy in the chambers. 
Two O-rings in grooves of the PBTP insulator seal the endplugs gas-thight inside the tubes. 
\begin{table}[htb]
    \centering
    \caption{sMDT drift tube design 
    parameters~\cite{sMDT_ATLAS_parameter_book}}
    \label{tab:Tube_Design_Parameters}
    \begin{tabular}{|c|c|}
      \hline
  Design parameter & Value\\
      \hline
  Tube material                             & Aluminium                \\
                                            & AW 6060-T6 / AlMgSi      \\
  Tube inner/outer surface                  & Surtec 650 chromatisation \\
  Tube inner diameter                       & $14.2 \pm 0.08$~mm       \\
  Tube outer diameter                       & $15 \pm 0.1$~mm          \\
  Tube wall thickness                       & $0.4 \pm 0.13$~mm        \\
  Tube length                               & $1615 \pm 0.2$~mm        \\
  Tube straightness deviation               & $< 0.5$~mm               \\
  Tube inner/outer wall coaxiality          & 0.1~mm \\
  Inactive tube length                      & 71~mm \\
      \hline
  Wire material                             & W-Re (97:3) \\
  Wire diameter                             & 50~$\mu$m \\
  with gold plating thickness               & 3~$\%$ \\
  Wire tension                              & $350^{+20}_{-15}$~g \\
  Wire resistance                           & 44~$\Omega$/m \\
      \hline
  Gas mixture                               & Ar:CO$_{2}$ (93:7) \\
  Gas pressure                              & 3~bar (abs.) \\
  Gas gain                                  & $2 \cdot 10^{4}$ \\
  Wire potential                            & 2730~V \\
  Maximum drift time                        & 175~ns \\
      \hline
    \end{tabular}
\end{table}
During the installation of the gas system, brass signal-caps with  gold-plated signal pins connecting to the high-voltage and readout distribution boards are screwed on the brass inserts of the endplugs, sealing with two O-rings the PBTP gas connectors between drift tubes and the gas gas distribution bars of each multilayer.
Gold-plated ground pins connect the grounding screws between the tube walls (see Section~\ref{sec:Chambers:Assembly}) to the ground of the front-end electronics.
The design and components of the sMDT endplugs are shown in 
Figure~\ref{fig:EndplugDesign}. 
Before the assembly of the drift tubes, all components are carefully cleaned. The raw aluminium tubes are vacuumed to remove any dust on the in- and outside. The endplugs and their components are cleaned in two steps in an ultrasonic bath for 15 minutes, first with an alkaline solution (Tickopur R30) at 50~°C and then with distilled water at 50~°C. All components are visually inspected before assembly in order rejecting incompletely deburred or damaged endplugs and tubes.

\begin{figure}[htb]
    \centering
    \includegraphics[width=0.95\textwidth]{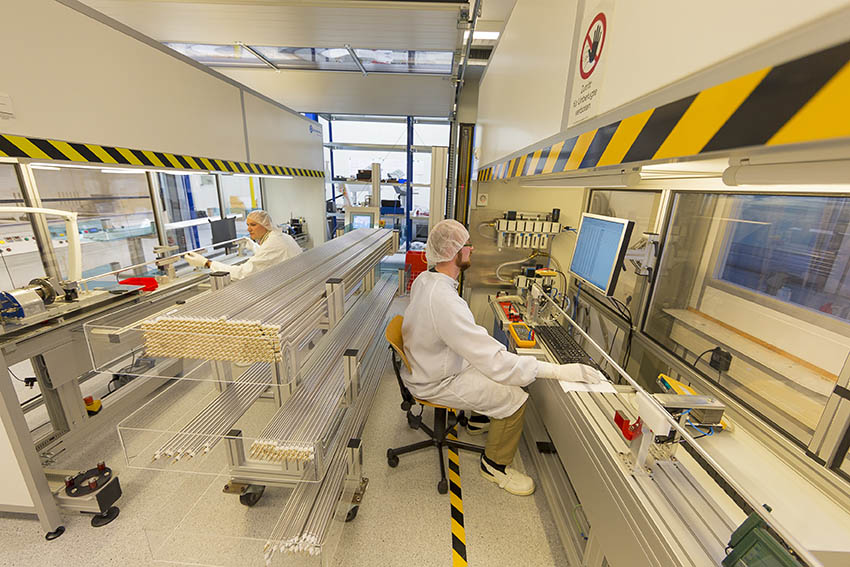}
    \caption{Clean room used for the drift tube production. The wire insertion station can be seen on the left, the wire tensioning station on the right.} 
    \label{fig:Wiring}
\end{figure}

Semi-automated processes are used for the drift tube assembly, 
performed at two separate stations for wire and endplug insertion and fixation and for wire tensioning and fixation under class 1000 flow boxes in a temperature and humidity controlled clean room (see Figure~\ref{fig:Wiring}, with the wire insertion and endplug swaging station to the left and the wire tensioning station to the right). 
The semi-automated processes ensure uniform production quality and avoid any touching of the sense wire length inside the tube.

At the wire insertion station, the sense wire is pulled from a spool by three motorised guiding wheels and fed into the tube through the inserted endplug by clean air flow at 6~bar pressure. By vacuum suction on the endplug on the opposite end of the tube, the wire is fed through the spiral wire locator and the brass insert. In order to fix the opposite endplug inside the tube and seal it gas tight, the tube wall is mechanically swaged on the O-rings of the endplug using a turning device. Air pressure is released, and the endplug on the side of the spool is swaged. 
On this end of the tube, the wire is crimped in a copper tubelet inserted in the brass insert of the endplug using a pneumatically operated crimping jaw which compresses the copper tubelet by 30~\%. 
At the wire tensioning station the sense wire is pulled by an actuator (linear motor) controlled by a force sensor on the tube end where the wire is not yet crimped. The wire is first overtensioned to 400~g for 10~s to reduce the later relaxation at the nominal tension of $350^{+20}_{-15}$~g. The tension is then released, and the wire is tensioned again to $360$--$370$~g and fixed by crimping on the free end completing the drift tube assembly. Integrated in the tensioning station is a commercial (CAEN SY502) electronic tension meter and a U-shaped permanent magnet which is raised from its parking position to surround the wired tube for the tension measurement via the frequency of wire oscillations excited in the magnetic field by injecting current pulses (see~\cite{ATLAS-TDR-10}).

The quality of the assembled drift tubes is ensured before their assembly in a chamber in a series of tests  (see Section~\ref{sec:QAQC:Tubes}). Every tube is labeled with a unique bar code number by which it is identifiable in the production database throughout the tests and later during chamber assembly. 
Production rates of more than 65 tubes per working day have been achieved by two technicians handling the tubes, operating the assembly stations and performing the tests in the same clean room.
\subsection{Chamber Assembly}
\label{sec:Chambers:Assembly}
The BIS1 chambers consist of 70 tubes per layer, while the BIS2-6 chambers have 58 tubes per layer. The chamber coordinate system is right-handed with 
the $x$ axis oriented along the tubes from the readout (RO) to the high-voltage distribution (HV) side, the $y$ direction is perpendicular to the chamber surface pointing from bottom to top in Figure~\ref{fig:sMDT_chamber_CAD}, i.e.\ outward in radius in ATLAS, and thus the $z$ axis points perpendicular to the tubes in the chamber plane, i.e.\ in beam direction towards the interaction point (IP). The side of the chamber oriented towards the interaction point in $z$-direction is denoted as IP side indicated in Figure~\ref{fig:sMDT_chamber_CAD}. The tubes within each layer are numbered starting from the IP side. The multilayers and the layers within each multilayer are numbered from bottom up in Figure~\ref{fig:sMDT_chamber_CAD}. 

\subsubsection{Precision Assembly Jigging}
\begin{figure}[htb]
  \includegraphics[width=1.0\textwidth]{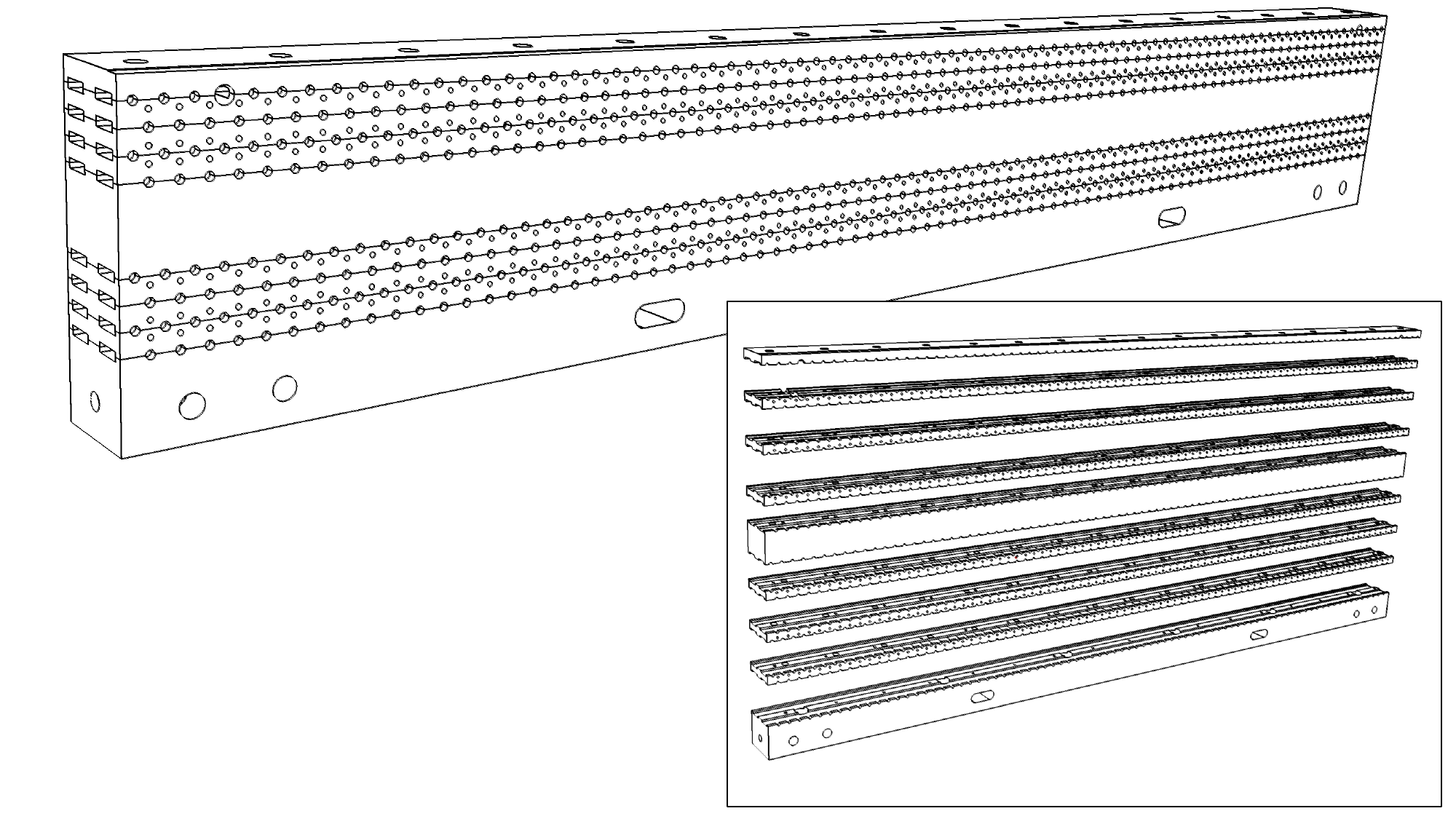}
  \vskip -38mm
    \includegraphics[width=0.45\textwidth]{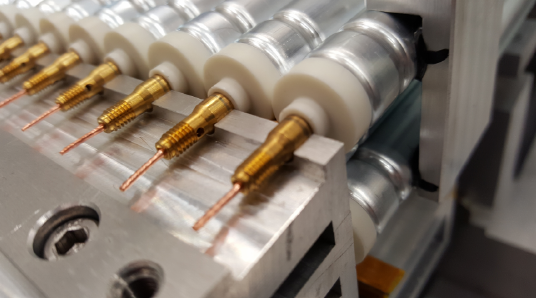}\hfill 
  \caption{The design of the precision assembly combs with illustration of the layered structure for the placement of the reference surfaces of the endplugs of the drift tubes in the fitting boreholes between the comb layers. The combs were machined in one piece, then the boreholes drilled with high precision following the desired wire grid, and finally the combs cut horizontally through the centers of the boreholes of each layer. The individual pieces can be screwed together again during chamber assembly. The insertion of the endplug reference surfaces in the half-cylindrical grooves of the combs for each layer is also shown.}
  \label{fig:assemblyCombs}
\end{figure}

\begin{figure}[htb]
  \centering
  \includegraphics[width=0.95\textwidth]{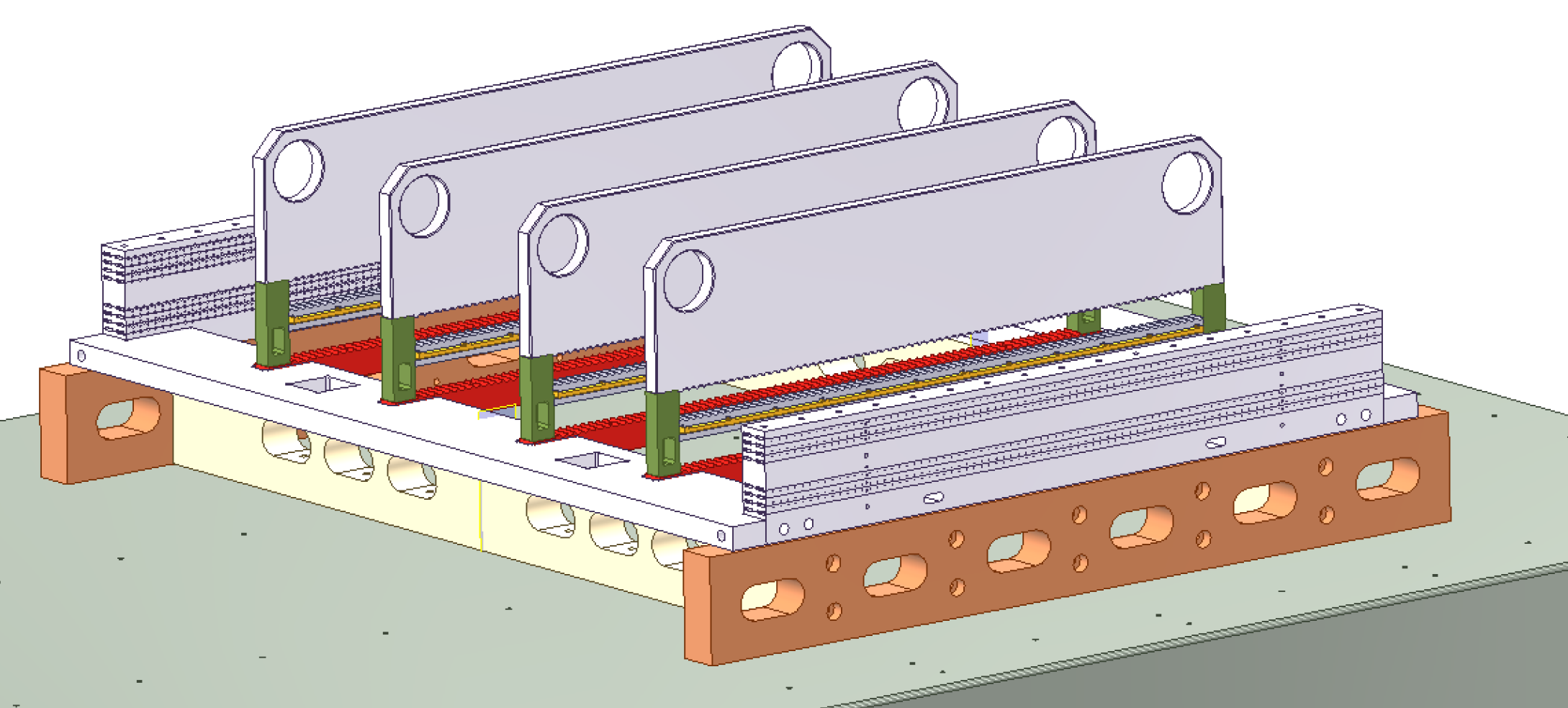}
  \caption{Chamber assembly jigging on a polished granite surface on which precise granite bars are mounted holding and aligning the assembly combs. 
  For guiding the tubes between the combs, four combs for the tube walls of the first layer are installed, and four weights are placed at the same longitudinal positions on top of each additional glued tube layer.}
  \label{fig:gluingsetup}
\end{figure}

The drift tubes are assembled int a chamber in a temperature- and humidity-controlled clean room nof class 10000.
A pair of aluminum combs with precise bore holes for insertion of the cylindrical endplug reference surfaces and segmented to allow for stacking and gluing of the tubes layers (see Figure~\ref{fig:assemblyCombs}) is used to position the drift tubes at their ends (see Figure~\ref{fig:gluingsetup}), thus defining the positions of the sense wires. After drilling of the bore holes, the combs have been cut horizontally through their centers leaving half-cylinders for placing and gluing the drift tubes of the next layer before screwing the next comb layer on top. 
The bore holes have a diameter of 5~mm fitting the external reference surfaces of the endplugs (see Figure~\ref{fig:TubeDesign}). Additional smaller holes with a diameter of 3~mm are used to insert the grounding screws in the gaps between adjacent drift tube layers after curing of the glue between them.
The two combs are aligned parallel to each other and perpendicular to the granite table in three dimensions via precise granite bars on which they are mounted (see Figure~\ref{fig:gluingsetup}). 
The centers and diameters of the boreholes were measured with respect to mounting reference surfaces on the combs with a coordinate measuring machine (CMM) and fitted to the wire grid with $z$ and $y$ pitch and multilayer $z$ and $y$ shifts and with individual layer distances for fixed $z$ pitch as free parameters. The results are shown in Table~\ref{tab:CMM_combs}.
The accuracy of the radial positions of the individual bore hole centers was determined to be 3~$\mu$m on the HV side and 4~$\mu$m on the RO side.

From the fabrication tolerance in the perpendicularity of the combs of 60~mm width to the table and of the coaxiality of the bore holes to the mounting reference surfaces of the combs of $\leq 0.01$~mm, the maximum angular deviation of endplugs is 0.5~mrad, leading to an additional wire displacement uncertainty at the end of the wire locator, which is 24~mm away from the endplug reference surface (see 
Figure~\ref{fig:TubeDesign}), of 4~$\mu$m, i.e. similar to the uncertainty in the bore hole positions. The aluminum tubes in their dense package with $100~\mu$m nominal distance between adjacent tube walls can have a maximum deflection during gluing in the chamber of $100~\mu$m over the first 30~cm from the combs, i.e.\ of 0.3~mrad, which gives an even tighter constraint. Measurements of wire positions in a prototype chamber with a muon beam at CERN~\cite{Bittner_thesis} showed no effect of potential angular deflections of the tubes on the sense wire positions.  
\begin{table}[htb]
    \centering
    \caption{Parameters of the combs used for BIS1-6 sMDT chamber assembly at MPI Munich measured with a coordinate measuring machine with order $1~\mu$m point resolution, compared to the design parameters of the dense wire grid. All values are in units of mm.}
    \label{tab:CMM_combs}
    \begin{tabular}{|l|l|l|r|}
      \hline
  Parameter & Value HV-side comb & Value RO-side comb & Design value \\
      \hline
 	Pitch $\Delta y$				 & $13.0762\pm0.00007$ & $13.0755\pm0.00007$ & 13.077 \\
 	Pitch $\Delta z$ 					 & $15.0997\pm0.00003$ & $15.1001\pm0.00003$ & 15.100 \\
    Multilayer distance $\Delta y_m$     & $45.6098\pm0.0008$ & $45.5894\pm0.0008$ & 45.600 \\    
 	Multilayer shift $\Delta z_m$		 & $-0.0013\pm0.0001$ &            $+0.0028\pm0.0002$ & 0.000 \\
 	distance layer $1\rightarrow 2$		 & $13.0794\pm0.0006$ & $13.0699\pm0.0006$ & 13.077 \\
 	distance layer $2\rightarrow 3$		 & $13.0736\pm0.0008$ & $13.0700\pm0.0008$ & 13.077 \\
 	distance layer $3\rightarrow 4$		 & $13.0762\pm0.0008$ & $13.0781\pm0.0008$ & 13.077 \\
 	distance layer $4\rightarrow 5$		 & $45.6098\pm0.0008$ & $45.5894\pm0.0008$ & 45.600 \\
 	distance layer $5\rightarrow 6$		 & $13.0754\pm0.0008$ & $13.0759\pm0.0009$ & 13.077 \\
 	distance layer $6\rightarrow 7$		 & $13.0738\pm0.0008$ & $13.0772\pm0.0009$ & 13.077 \\
 	distance layer $7\rightarrow 8$		 & $13.0766\pm0.0008$ & $13.0762\pm0.0009$ & 13.077 \\
      \hline
    \end{tabular}
\end{table}
During the assembly and curing of the glue, the tube walls are aligned with less precise combs from the bottom for the first layer of each multilayer and from top of each glued layer with weights (see Figure~\ref{fig:gluingsetup}). 
The minimum distance between the tubes is kept at longitudinal positions between the weights by $50\mu$m thick foils wrapped around the tubes.  
\begin{figure}[htb]
  \centering
  \includegraphics[width=0.8\textwidth]{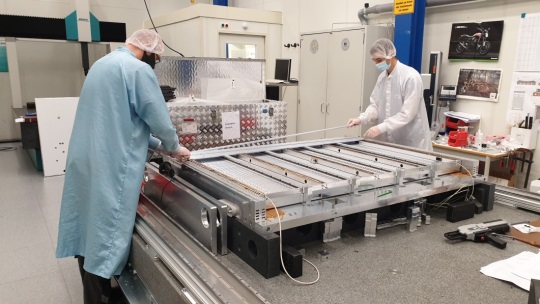}
  \caption{Assembly of an BIS1-6 sMDT chamber in the temperature controlled clean room. Two technicians are needed for the handling of the drift tubes and their insertion into the combs. The glue is distributed by an automated gluing machine installed on the granite assembly table.}
  \label{fig:chamber_assembly}
\end{figure}

\subsubsection{Chamber Gluing Procedure}
Chambers are assembled layer by layer. A spacer structure, described in more detail in Section~\ref{sec:Chambers:Assembly:IPA}, is glued between layer~4 and 5. The drift tubes of each layer are placed on the stacked combs. 
For each tube the passing of the quality criteria (see Section~\ref{sec:QAQC:Tubes}) is automatically verified from the production database by reading the bar code, entering at the same time the tube position in the chamber in the database. 
The tubes are glued to the neighbouring tubes of the preceding layers using two-component epoxy resin (Araldite 2011) 
distributed in six equidistant 5~cm long stripes of about 0.5~ml along the tubes of the preceding layer with an automated glue dispenser. 
Then the tubes of the new layer are placed in the assembly combs and the next comb layer is screwed to the previous one. In particular, the ends of the plastic insulators of the endplugs sticking out of the tubes are glued together to ensure stable wire positioning after releasing the chamber from the assembly table. 

\begin{figure}[htb]
	\centering
  \subfloat[]{\includegraphics[width=0.55\textwidth]{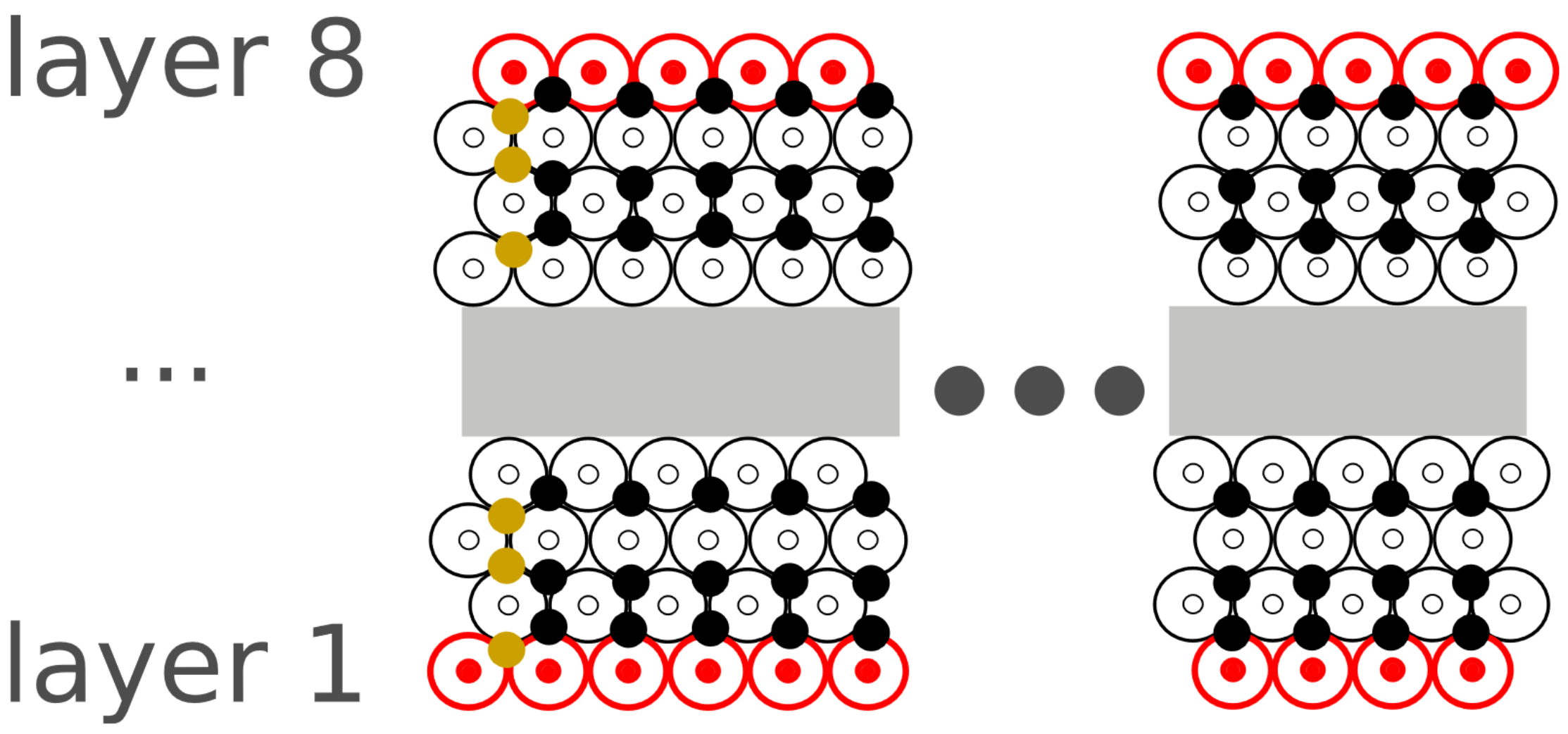}\label{Fig:groundingScrewScheme}}
	\subfloat[]{\includegraphics[width=0.44\textwidth]{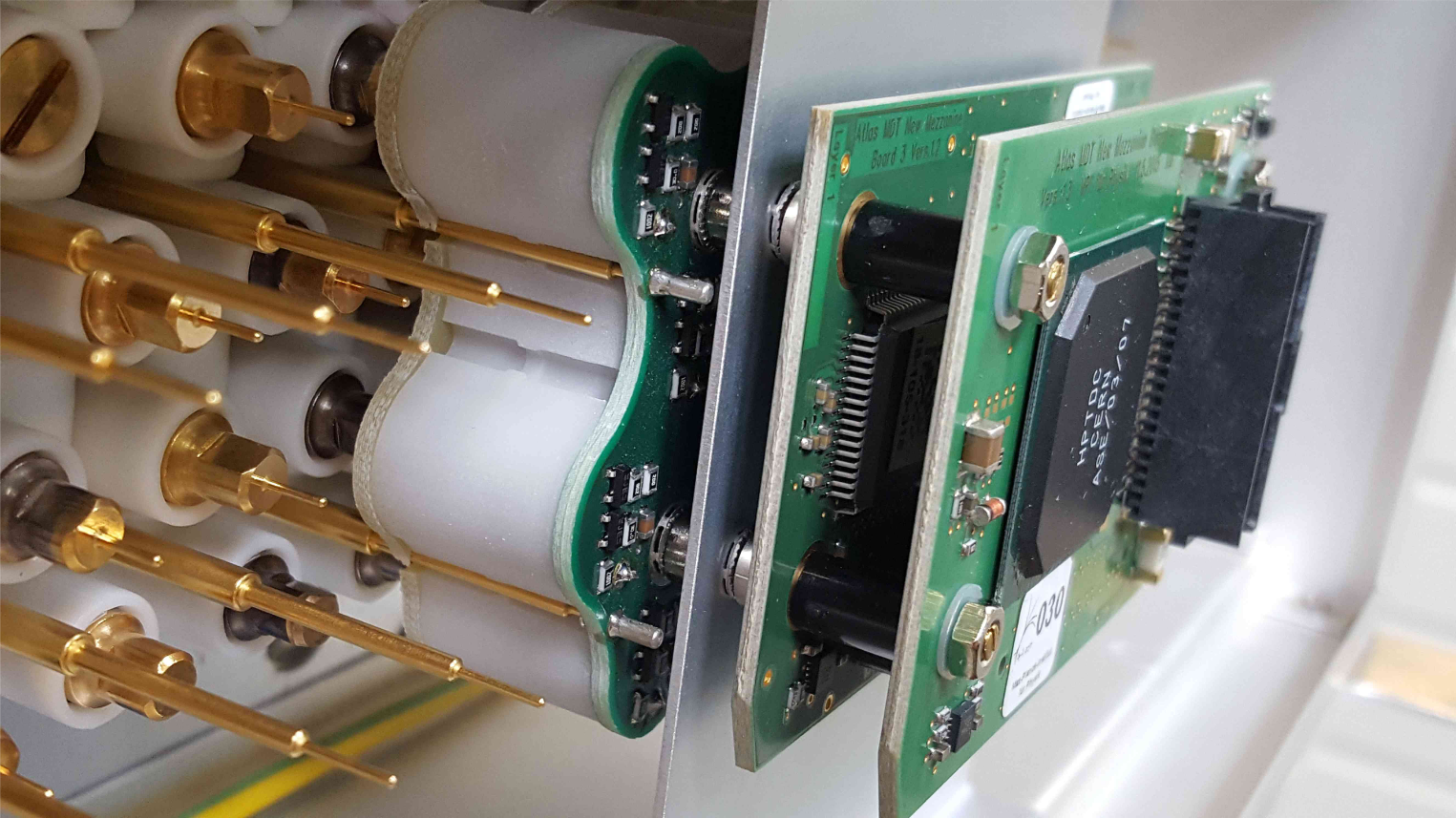}\label{Fig:grounding_mezz}}    
	\caption{a) Locations of the grounding screws between adjacent tubes providing electrical ground contacts to the drift tube walls (see also Figure~\ref{fig:EndplugDesign}). Black-marked screws are installed during the chamber gluing while the gold-marked screws are installed during the gas system installation at multilayer edges on one side. b) Gold plated ground pins screwed onto the grounding screws to connect to the high-voltage and signal distribution boards on the chambers which also connect to sense wires via the signal caps sealing the gas connectors to the endplugs. The latter are shown in the picture together with the active readout mezzanine cards connected to them.} 
	\label{Fig:groundingScrews}
\end{figure}

The grounding screws between adjacent tube layers are inserted through the corresponding holes in the combs after curing of the glue between the layers while the chamber is still in the assembly jigging.
The grounding screw grid is shown in Figure~\ref{Fig:groundingScrewScheme}). It ensures ground contacts between all adjacent tubes and avoids interference of the ground pins screwed later on the grounding screws (see Figure~\ref{Fig:grounding_mezz} with the gas connectors (see Section~\ref{sec:Chamber:Gas}). 
The glue is usually left for curing overnight with the weights placed on top of the chamber (see Figure~\ref{fig:gluingsetup}). 
However, the glue used becomes sufficiently solid after a curing time of four hours such that two layers of drift tubes can be glued on the same day saving two working days per chamber if needed. The gluing of the pre-assembled spacer frame on the first multilayer requires an extra day because of the necessary alignment on the table and the cabling and test of the in-plane monitoring system, in total five working days for the tube assembly in the accelerated scheme and seven without.  

\subsubsection{In-Plane Alignment System}
\label{sec:Chambers:Assembly:IPA}
\begin{figure}[htb]
	\centering
	\includegraphics[width=0.45\textwidth]{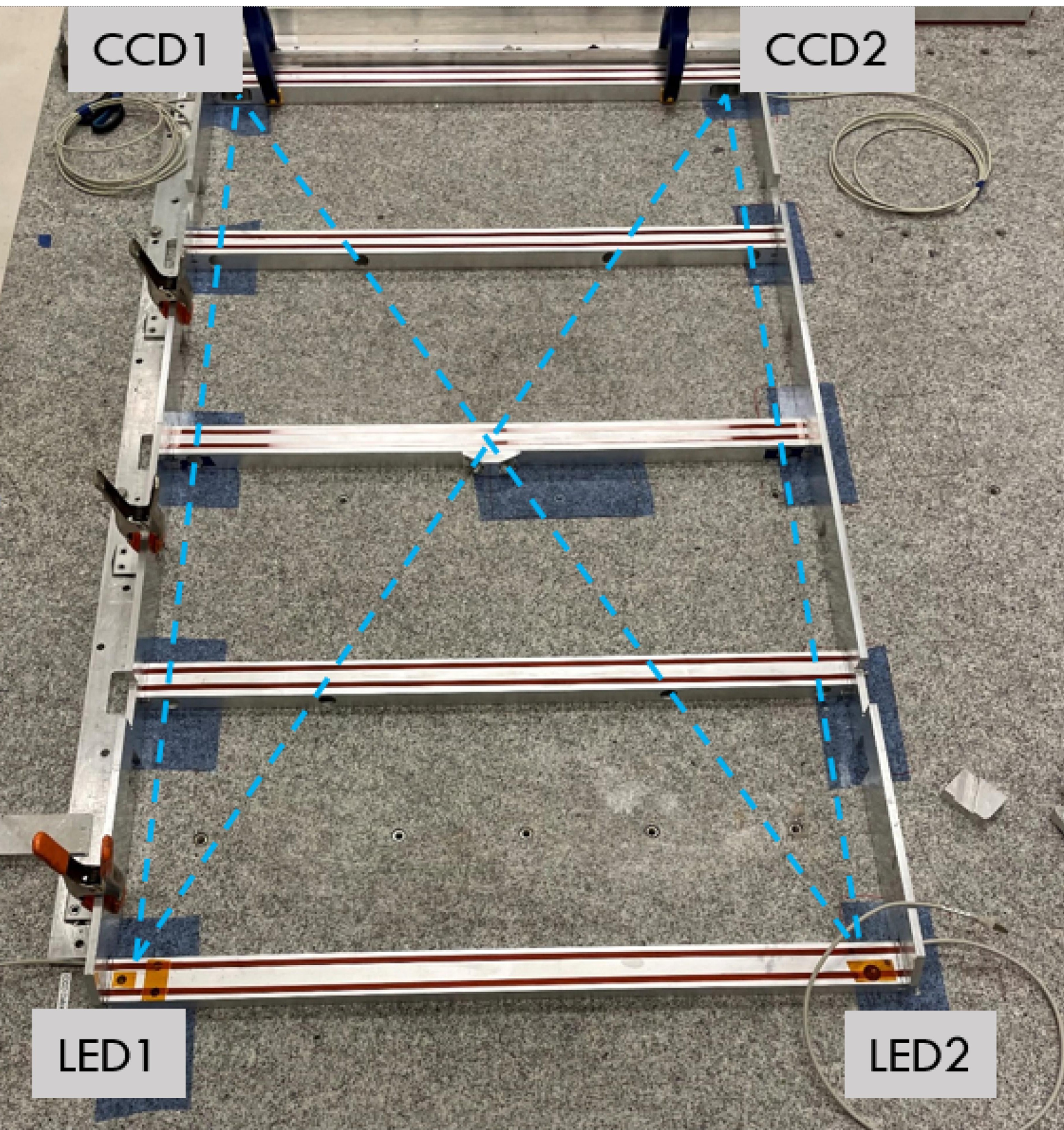}    
	\includegraphics[width=0.53\textwidth]{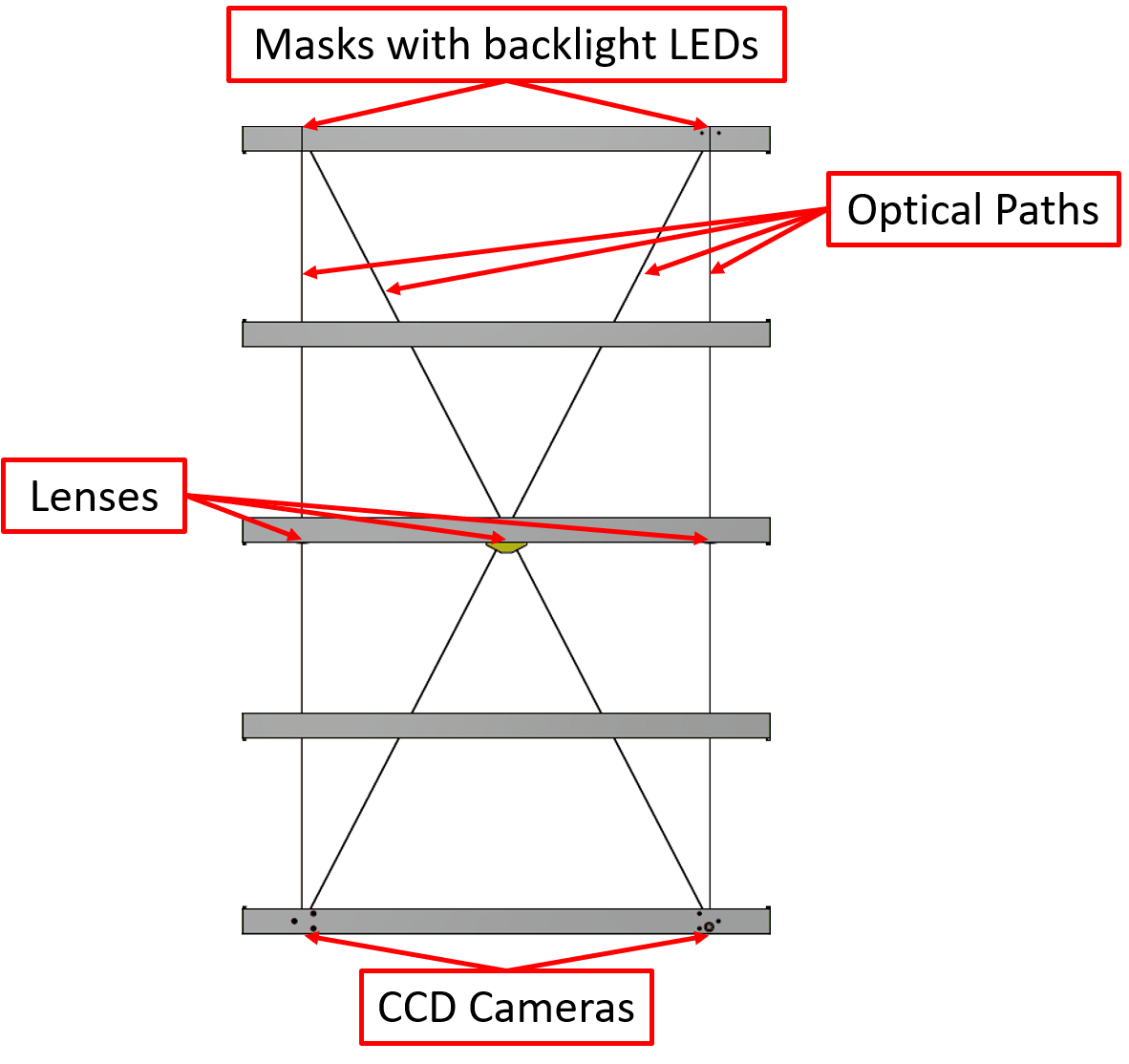}
	\caption{Spacer with longitudinal and diagonal RASNIK straightness sensors consisting of an LED illuminated mask, a lens and a pixel camera (CCD) pre-assembled and tested on a granite table before mounting on the chamber.}
	\label{Fig:spacer}
\end{figure}

The aluminum spacer frame glued between the two multilayers consists of five equidistant crossbars connected by two longitudinal beams. 
It houses the in-plane alignment system (IPA) which employs two longitudinal and two diagonal RASNIK  straightness sensors~\cite{Beker_2019} connecting two infrared LED illuminated coded masks and two pixel CCD cameras on the crossbars of the spacer on the HV and RO sides, respectively, via four lenses in the middle cross bar.
Figure~\ref{Fig:spacer} shows the optical paths of the IPA. 

The RASNIK sensors detect relative movements of mask, lens and camera with micrometer precision to monitor deformations of the chamber under gravity and external stress. The longitudinal sensors measure the gravitational sag of the chamber along the tubes which happens to be close to the wire sag such that no adjustments are needed. Most importantly, the diagonal sensors measure the torsion angle between the HV and RO sides of the chamber. Because of the small spacer height, the BIS1-6 sMDT chambers are vulnerable for torsion deformations under external forces or internal stresses. The \say{zero-readings} of the RASNIK sensors corresponding to perfect chamber planarity are taken during chamber assembly on the granite table and are stored in the chamber production database.

\subsubsection{Support Structures and Alignment Platforms}
The chamber support structures carrying the bearings (see the mounting points in Figure~\ref{fig:sMDT_chamber_CAD}) for the installation on the ATLAS rail system mounted on the toroid magnet coils 
and platforms for mounting of the optical sensors of the ATLAS global muon chamber alignment are glued on top of the second multilayer with two-component epoxy resin DP490 also used for gluing of the spacer.
Each chamber is equipped at the corners with four so-called Axial/Praxial (AP) platforms for sensors connecting neighbouring chambers within the same sector of the ATLAS muon spectrometer (see Figures~\ref{fig:sMDT_chamber_CAD} and \ref{fig:BIS_layer}. On  chambers in certain locations in the BIS layer, up to three platforms for chamber-to-chamber connections (CCC) between neighbouring sectors are distributed on the outer multilayer surface. In addition, two or four platforms for magnetic field sensors are glued to the same layer.

The platforms have precision reference surfaces for the mounting of the optical sensors later at CERN. Their positions have to be known with a precision of $10~\mu$m with respect to the wire grid in the $y$ and $z$ coordinates and with order $100~\mu$m acuracy along the tubes to allow for precise enough monitoring of the relative chamber positions and layers in ATLAS~\cite{ATLAS-TDR-10,ATLAS-TDR-26}.
Insulating tape 
is attached on the bottom of each platform before gluing to ensure that there is no electrical contact to the tubes. 

\begin{figure}[htb]
	\centering
	\subfloat[]{\includegraphics[width=0.38\textwidth]{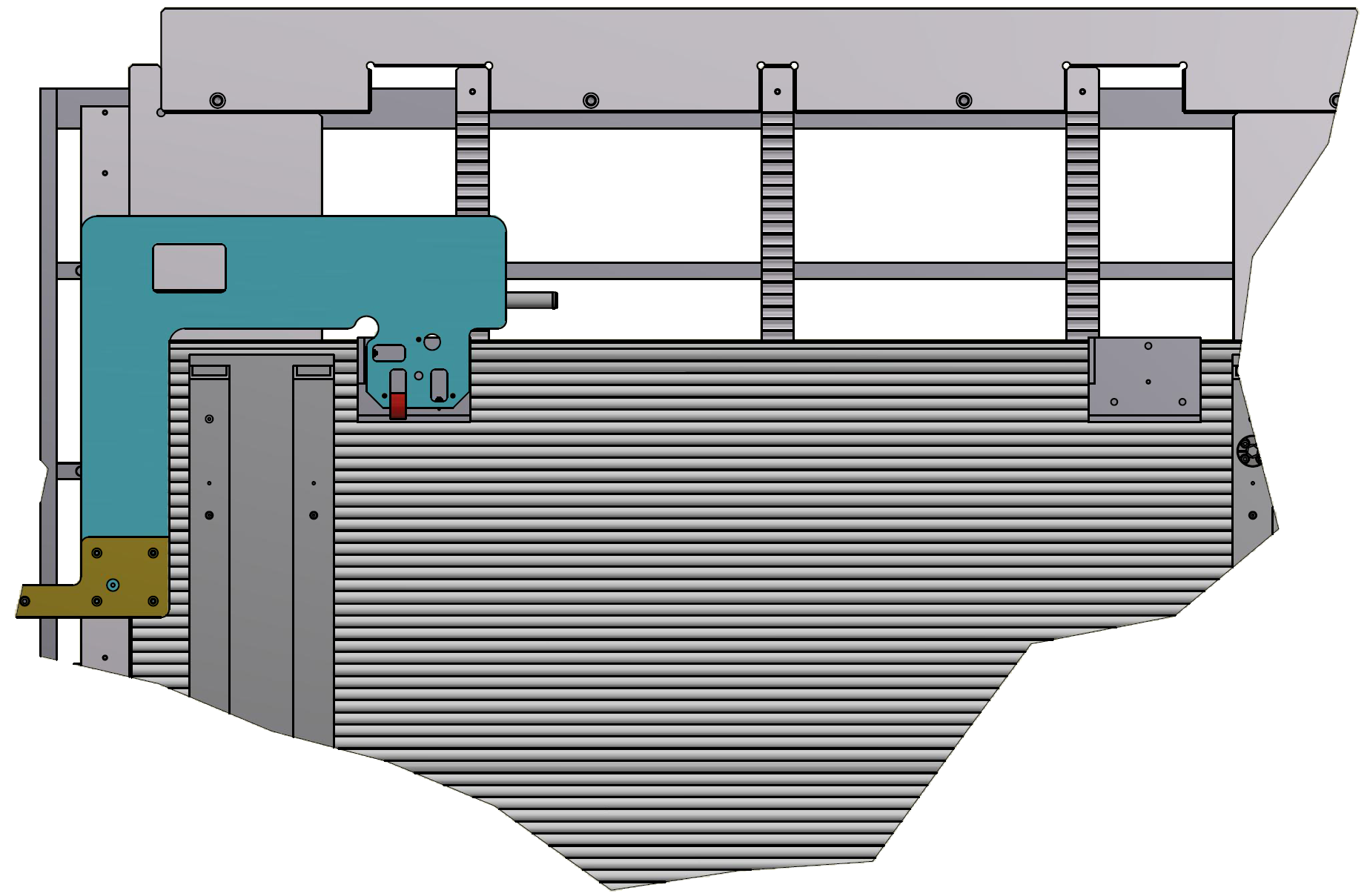}\label{Fig:Platform_gluing:AP}} 
	\subfloat[]{\includegraphics[width=0.59\textwidth]{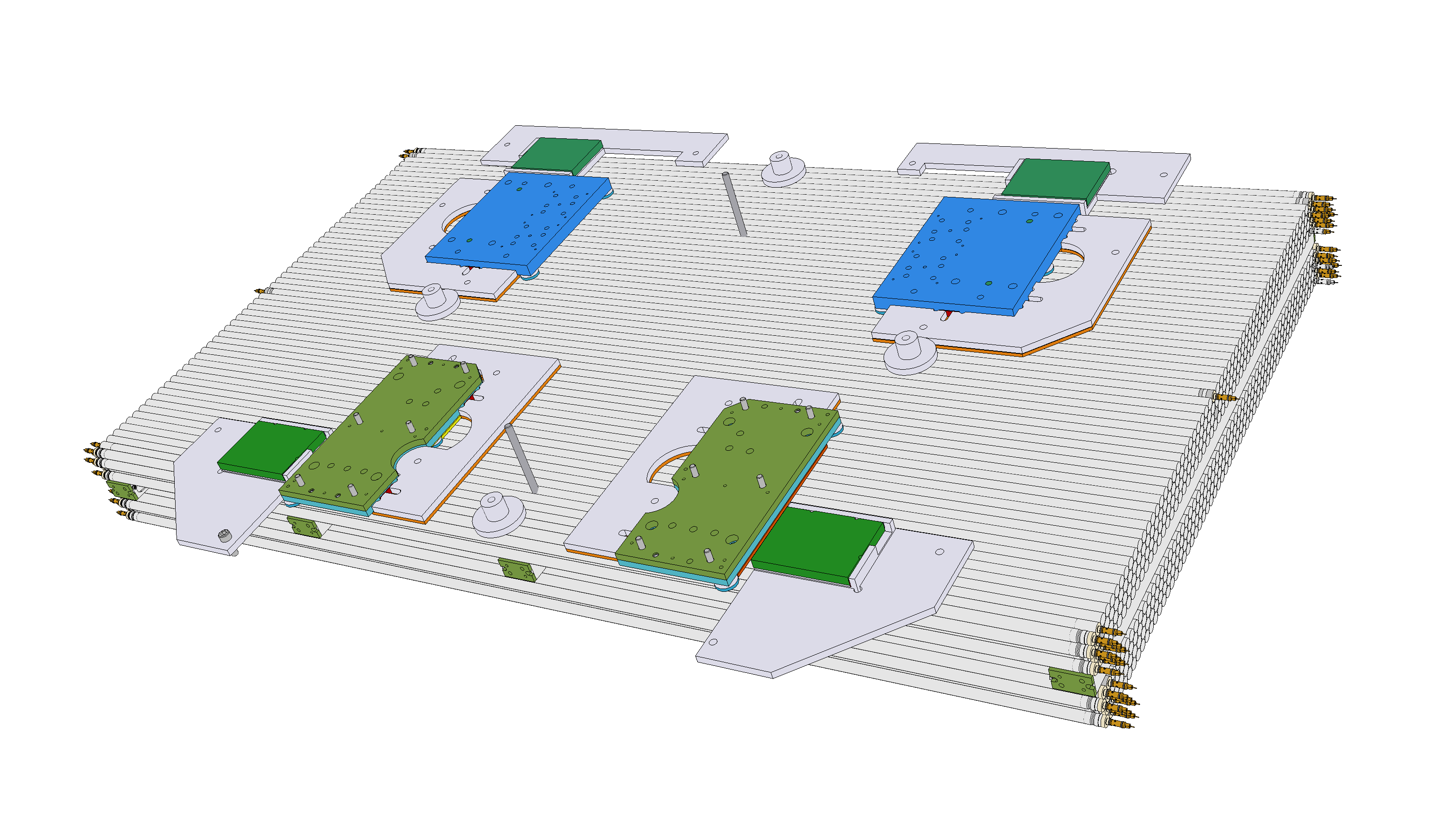}\label{Fig:Platform_gluing:CCC_B}}
	\caption{a) Precision mounting bracket for the AP platform at one corner of the chamber reaching around the already glued chamber support profile to the next comb. b) Brackets for the positioning of the CCC and magnetic field sensor platforms in different locations on the chamber surfaces relative to the AP platforms (shown on two corners at the upper side of the picture). These brackets are mounted with high precision, surveyed with the coordinate measuring machine, on a common plate (not shown) which is attached to the four AP platforms for simultaneous gluing to the chamber.}
	\label{Fig:Platform_gluing}
\end{figure}

First, the AP sensor platforms are positioned and glued to the tubes with respect to the assembly combs using four L-shaped brackets (see Figure~\ref{Fig:Platform_gluing}a). Each mounting bracket allows for precise 3D adjustment of the platforms they hold at six contact points on the reference surfaces, one in $x$, three in $y$ and two in $z$ direction. Before the gluing, the positions of the AP platforms are measured using an electro-mechanical 3D feeler arm to verify their position and orientation with respect to the combs and if necessary adjust them. The chamber support profiles are glued to the chamber simultaneously, also electrically insulated by tape.

After the curing of the glue over night, the platforms for the CCC and magnetic field sensors are positioned with respect to the glued AP platforms using a large mounting plate where the positions of the platforms are defined relative to four blocks matching the reference surfaces of the AP platforms (see Figure~\ref{Fig:Platform_gluing}b). Finally, all platform positions and angles are measured relative to the combs with the 3D feeler arm (see Section~\ref{sec:QAQC:Geometry:Platforms}. The gluing and measurement procedures of the external alignment platforms require two additional working days.
Before the installation of the gas distribution system and electronics, the positions of the sense wires are measured with the CMM in the clean room as long as the reference surfaces on the endplugs are still accessible (see Section~\ref{sec:QAQC:Geometry:WirePos}).

\subsubsection{Gas Distribution System}
\label{sec:Chamber:Gas}
\begin{figure}[htb]
\vskip -18mm
    \hskip -5mm\includegraphics[width=0.9\textwidth]{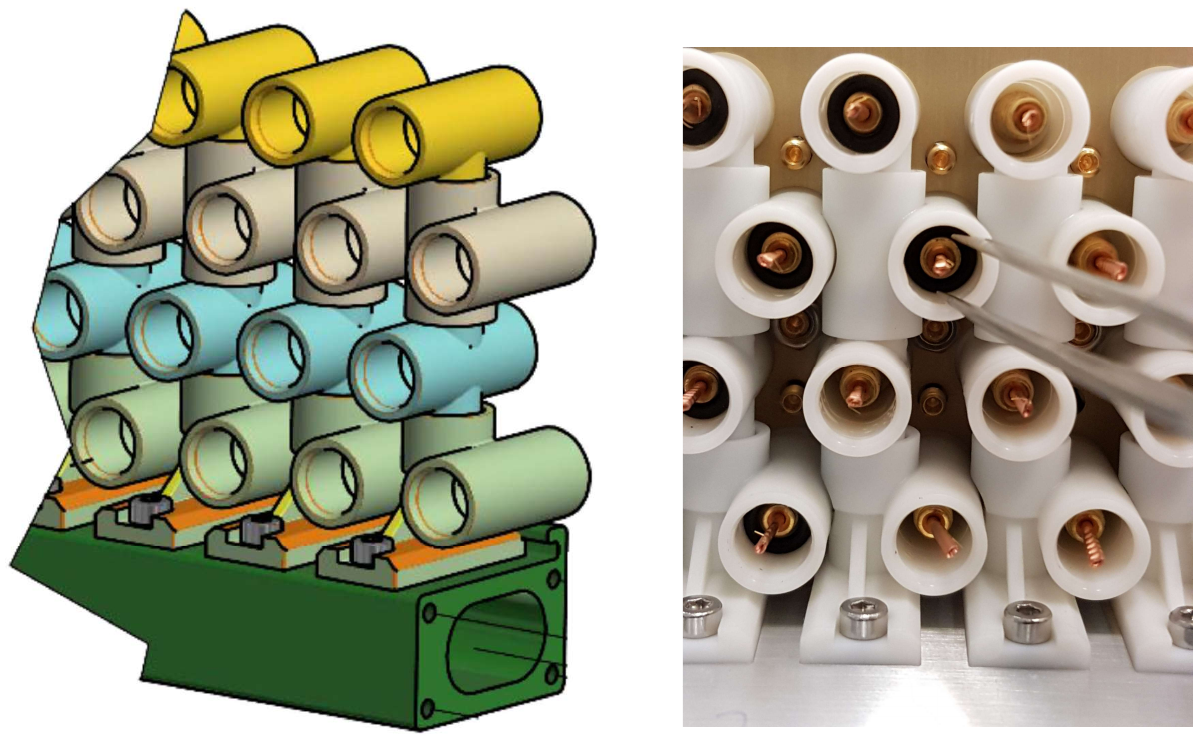}
	\vskip -22mm
    \caption{The array of injection molded plastic gas connectors linking columns of four drift tubes in the four tube layers of a multilayer serially to the common aluminum gas distribution bar at each chamber end (see also Figure~\ref{fig:EndplugDesign}).} 
	\label{Fig:gas_system}
\end{figure}

The gas distribution system consists of a modular array of injection molded plastic gas connectors of PBTP material without glass fiber on the endplugs, interconnected between the four neighbouring endplugs in the four layers of a multilayer. The components are cleaned like the endplugs. The connectors are sealed to the endplugs with the signal caps (see Figure~\ref{fig:EndplugDesign}). The columns of four gas connectors are attached to an aluminum gas distribution bar for each multilayer which is connected to a copper gas pipe terminated by a valve on the chamber (see Figure~\ref{Fig:gas_system}). The gas bars are chromatized on the in-and outside for cleaning and reliable ground contact. They are integrated in the Faraday cages for each multilayer on the RO and the HV side, and electrically insulated from the gas pipes routed on the chamber outside. Together with the two O-rings sealing the endplugs in the tubes, the BIS1 sMDT chambers contain 5600 O-rings seals.
First, the aluminum ground foils, the 0.5~mm thick flexible back planes of the Faraday cages with holes for the brass inserts, are mounted on the multilayers on both sides and fixed by nuts with washers on the grounding screws, accommodating slightly varying tube lengths (see Figure~\ref{fig:RO_electronics}).
Then the pre-assembled gas manifolds (see Figure~\ref{Fig:gas_system}) are installed on the multilayers on both sides in a temperature controlled clean room of class 10000, sealing the tubes with the signal caps screwed on the threads on the brass inserts with additional O-rings (see Figure~\ref{Fig:gas_system}, and the grounding pins are screwed on the ground screws between the gas connectors. 

After the gas system assembly, the gas leak rates of the multilayers are measured (see Section~\ref{sec:QAQC:Gas}) ensuring that they are below the required limit of $2 N_{\text{tubes}}\cdot 10^{-8}~\text{bar}\cdot\text{liter}\cdot\text{s}^{-1}$, where $N_{\text{tubes}}$ is the number of tubes in a multilayer. In case the measured leak rate exceeds the limit, leaks are searched for and repaired until the limit is retained. In spite of the many O-rings this procedure converged quickly within few hours for all chambers.
\begin{figure}[htb]
  \centering
  \includegraphics[width=0.9\textwidth]{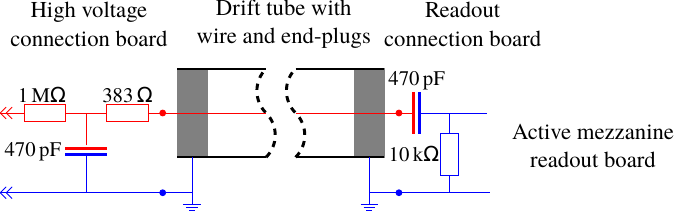}
  \caption{Passive electronics circuits for each drift tube on the signal and high-voltage distribution boards connected to the sMDT multilayers.}
  \label{fig:passiveEl}
\end{figure}
\begin{figure}[htb]  
  \centering
  \includegraphics[width=0.90\textwidth]{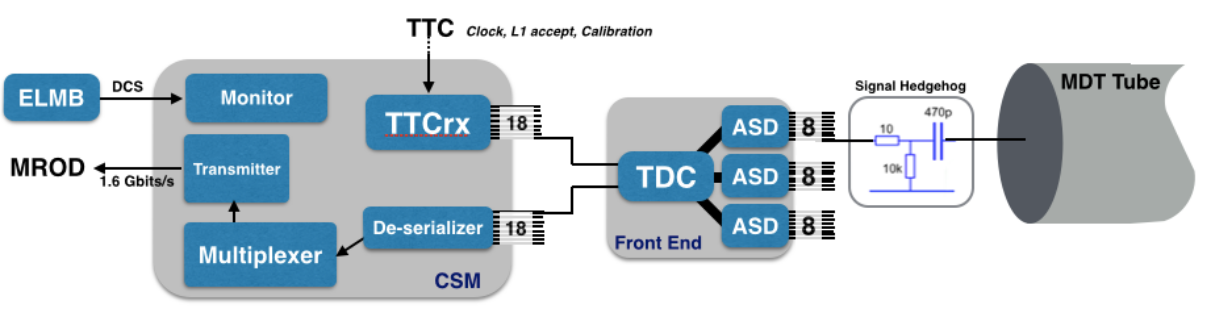}
  \caption{Block diagram of the sMDT readout electronics chain.}
  \label{fig:feChain}
\end{figure}
\begin{figure}[htb]
  \centering
  \includegraphics[width=0.53\textwidth]{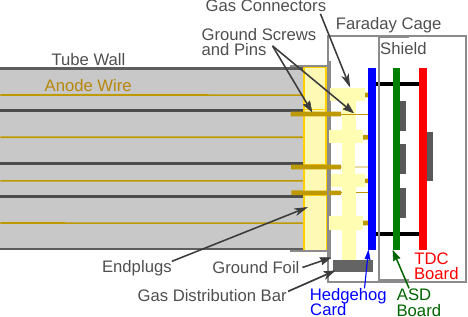}
  \includegraphics[width=0.43\textwidth]{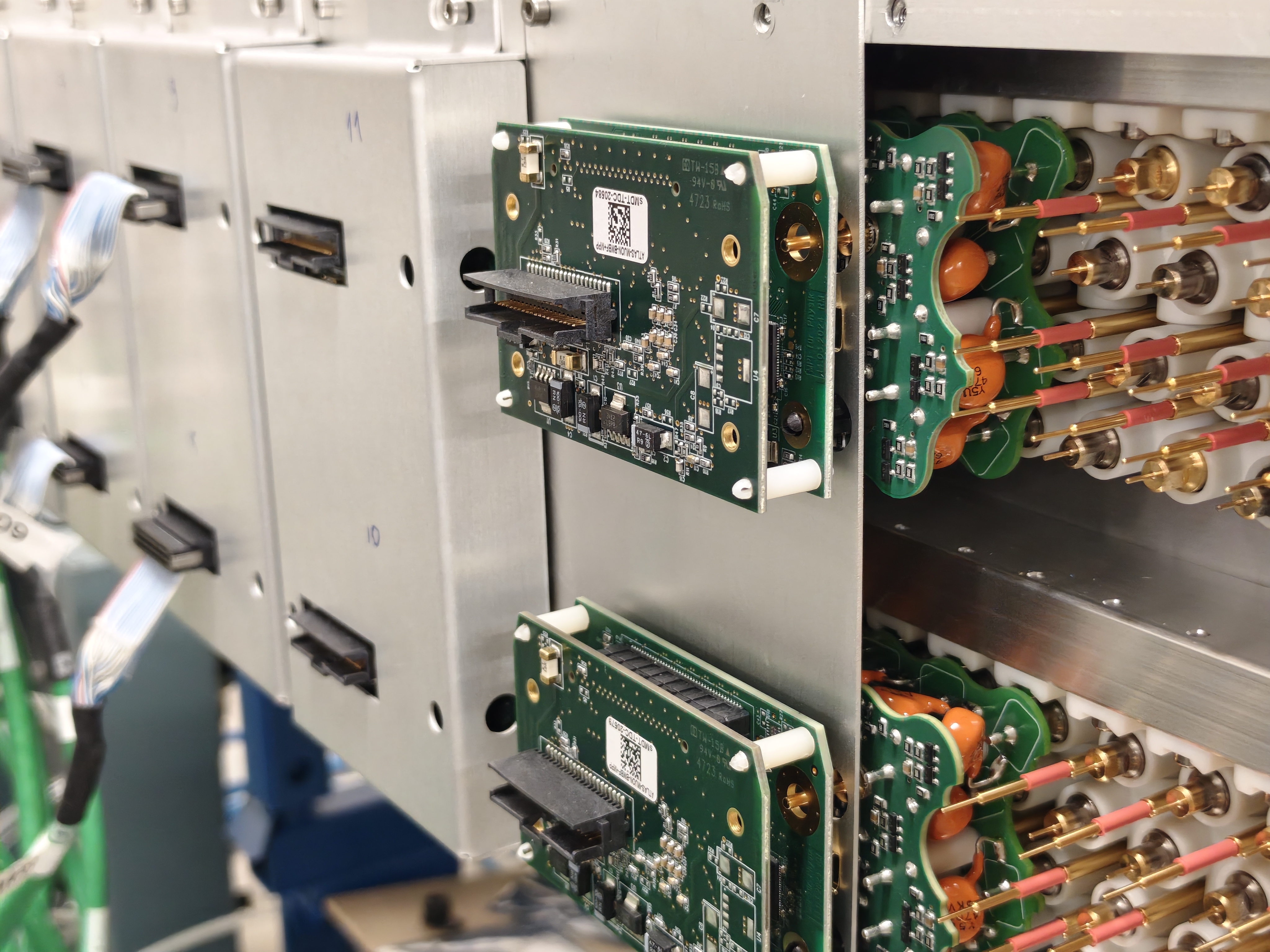}
  \caption{Layout of the RO-side Faraday cage of each multilayer with the 
  24-channel signal distribution (hedgehog) boards and the stacked mezzanine cards, consisting of an ASD board with three 8-channel ASD chips and a board with the TDC chip (see also Figure~\ref{fig:feChain}), shielded from the hedgehog boards by an aluminum cover plate. The backplane of the Faraday cage consists of a 0.5~mm thick chromatized aluminum ground foil connected to the ground pins with nuts. The mezzanine cards are enclosed in individual aluminum boxes. The gas distribution bars are also incorporated in the Faraday cage, electrically insulated from the external gas pipes.}
  \label{fig:RO_electronics}
\end{figure}

\subsubsection{High-Voltage Distribution and Readout Electronics}
\label{sec:Chamber:Electronics}
After installation of the HV- and RO-side Faraday cages made of 1~mm thick chromatized aluminum sheets (see Figure~\ref{fig:RO_electronics}),  
the high-voltage distribution and the readout electronics boards (so- called hedgehog cards for historical reasons) are mounted onto the signal and ground pins (see Figure~\ref{Fig:grounding_mezz}).
The boards serve $6\times 4$ drift tubes in a multilayer at both edges  of the multilayers $5\times4$ tubes. 
The passive electronics circuits on the hedgehog cards are illustrated in Figure~\ref{fig:passiveEl}. The coupling capacitors are enclosed in plastic cylinders sandwiched between printed-circuit boards to prevent dark currents and discharges on the hedgehog cards under HV in the dense environment (see \ref{Fig:grounding_mezz}). 

Each multilayer is supplied independently with high voltage.
Each HV supply line is further split into four lines, one for each tube layer, which can be disconnected independently, in HV distribution boxes mounted on the HV Faraday cages, and are routed inside the cages to the HV hedgehog cards which are daisy-chained in each tube layer. During the installation of the electronics 12 temperature sensores are installed on each chamber on the top and bottom multilayer. 

The traces for the unamplified signals on the RO hedgehog cards are shielded against pick-up noise from the active RO mezzanine cards by the cover plates of the Faraday cages (see Figure~\ref{fig:RO_electronics}
A stacked architecture is used for the active RO electronics on the signal distribution board as illustrated in Figure~\ref{fig:RO_electronics}. The mezzanine cards consist of stacked boards with three 
Amplifier-Shaper-Discriminator (ASD) 
chips~\cite{ASD2_IEEESensor,ABOVYAN2019374,ASD2_manual} with eight channels each and a board carrying the TDC ASIC~\cite{GUO2021164896}. Both chips have been developed for the readout of the (s)MDT chambers at the HL-LHC supporting the increased data rates and continuous readout for the (s)MDT based first-level muon trigger~\cite{ATLAS-TDR-26}. 

The ASD chip contains an 8-bit Wilkinson ADC which measures the charge in the leadings edge of the shaped signal as a measure of the signal amplitude for time-slewing corrections. It introduced an additional deadtime which is programmable between 180~ns and 720~ns, around the maximum drift times of the sMDT and MDT tubes, respectively. For the sMDT chambers always the minimum deadtime is used. The TDC chips pass the digitized timing information of the 24 channels of each mezzanine card via shielded cables to the chamber service modules (CSM) which can serve up to 20 mezzanine cards and forward the multiplexed data via optical fiber cables to the DAQ system and muon trigger processors. The BIS1 chambers are read out by two CSMs, the BIS2-6 chambers one which for space reasons are mounted in ATLAS off-chamber on the toroid magnet coils. The (s)MDT chamber readout electronics chain is shown in Figure~\ref{fig:feChain}.

\section{Quality Control Procedures}
\label{sec:QAQC}
The drift tube production and chamber construction are subject to strict requirements on quality and performance. Quality control test have been performed for every drift tube and chamber during the construction process. The performance of the completed sMDT chambers is verified by measurements of cosmic muon tracks.

A Web-based application was developed in order to automate the evaluation of the quality control measurements and to store the results in the production database~\cite{Rendel:2023,Buchin:2026}. 
It uses the following workflow. At defined steps in the production certain measurements are performed. 
The measurement data together are uploaded via the interface to the Webpage triggering an automated analysis of the measurement. The results are then displayed and stored in a MySQL database together with the raw data and additional information about the measurement. A notification mail is sent to all concerned persons. 
The Webpage allows graphical access to all measurements for a specific chamber as well as comparison plots over the whole series production. The application is shared between both production sites.

\subsection{Drift Tube Tests}
\label{sec:QAQC:Tubes}
The quality control acceptance criteria of the drift tubes for chamber assembly based are summarized in Table~\ref{tab:tube_limits}.
\begin{table}[htb]
    \caption{Requirements for drift tube quality control tests for BIS1-6 sMDT chamber construction}
    \centering
    \begin{tabular}{|l|c|}
  \hline
  Tube length &$1624.4 \pm 0.75$~mm\\
  Wire tension &$335--370$~g\\
  Tension loss after 2 weeks &$<18$~g\\
  Argon leak rate &$<10^{-5}~\text{mbar} \cdot \text {liter}/\text{s}$ \\
  Dark current &$<2$~nA\\
      \hline
    \end{tabular}
    \label{tab:tube_limits}
\end{table}
A total of 26805 drift tubes was produced between June 2019 and March 2023, of which 25098 were accepted for chamber assembly. Drift tubes were rejected either because their assembly failed or they did nopt pass one of the  quality control tests described in detail below. 23040 drift tubes were needed for the construction of the 48 
A-side BIS1-6 sMDT chambers. One BIS1-type and three BIS2-6-type spare chambers were constructed in addition, resulting in a total of 24992 drift tubes assembled in chambers. 
In total, 1650 drift tubes were rejected, corresponding to an overall rejection rate of $6.3\%$ (see Table~\ref{tab:rejection_Rates}). 
\begin{table}[htb]
    \caption{Rejection rates of drift tubes in the different quality control steps during serial production}
    \centering
    \begin{tabular}{|m{4.5cm}|c|}
      \hline
  Production step & Rejection rate [$\%$]\\
      \hline
  Assembly                                            & 2.0 \\
  Tube length meas.                                   & 0.1 \\
  First wire tension meas.                            & 1.1 \\
  Last wire tension meas. and Wire tension loss       & 2.0 \\
  Gas leakage meas.                                   & 0.02 \\
  HV leakage current meas.                            & 1.1 \\
      \hline
  Total                                               & 6.3 \\
      \hline
    \end{tabular}
    \label{tab:rejection_Rates}
\end{table}

\subsubsection{Tube Length Measurement}
Due to mechanical tolerances of the raw tubes and in the drift tube assembly procedure, the assembled drift tube lengths can vary by up to 1~mm. Large variations in tube lengths within one multilayer could cause difficulties in the installation of the gas manifolds and of the ground foils. Therefore, lengths of the assembled drift tubes were measured using a feeler gauge, and the drift tubes assigned to one of three length categories. Only drift tubes belonging to the same length category, with lengths varying by at most 0.5~mm, are assembled in a common multilayer. 

\subsubsection{Wire Tension Measurement}
The sense wire positions need to be known with high precision not only at the tube ends but along the whole tube. In the region between the wire locators, the wire position is determined by the gravitational sag of the wire which in turn depends on the wire tension~\cite{ATLAS-TDR-10}. The wire tension is required to be between 335 and 370~g. It is measured with the wire tension meter (see Section~\ref{sec:Chambers:Tubes}) immediately after the assembly to verifying the tension achieved. 
A second wire tension measurement is performed after two weeks in which the main relaxation of the tension takes place and then stabilizes. On average, the tension decreases by about $10$~g.
A third tension measurement is performed as close as possible to the gluing of the chamber to ensure that the wire has been reliably crimped and did not slip over time. Only drift tubes with final wire tension between 335 and 370~g and a tension loss below 18~g between first and last measurement are accepted for chamber assembly.

\subsubsection{Gas Leak Tightness Certification}
The gas leakage rate of the sMDT chambers is required to be below $10^{-5}~\text{mbar}\cdot\text{liter}/\text{s}$. For the leak rate measurement, the drift tubes are enclosed in a stainless steal cylinder while they are sealed with a signal cap on one end and connected to a gas supply with sMDT gas connector like on the chambers on the other. The test cylinder is connected to a helium leak detector (Leybold Phoenix L300i) including a vacuum pump. The drift tube is first evacuated and then filled with a gas mixture of $95\%$ argon and $5\%$ helium at an absolute pressure of 2290 mbar which corresponds to $10\%$ higher overpressure with respect to the vacuum in the test cylinder than the overpressure with respect to  atmospheric pressure of the drift tubes at the nominal operating pressure of 3~bar absolute. The cylinder is then evacuated, and the helium leak rate from the drift tube into the vacuum measured by means of the mass spectrometer of the leak detector.

For a drift tube to be accepted, the argon leak rate, inferred from the helium leak rate, has to stabilize below the specified leak rate limit
after evacuation of the cylinder. The argon leak rate sensitivity of the setup was 
$3\cdot 10^{-8}~\text{mbar} \cdot \text{liter}/\text{s}$ corresponding to a detectable helium leak rate of $5\cdot 10^{-9}~\text{mbar} \cdot \text{liter}/\text{s}$.

\subsubsection{High-Voltage Leakage Current Test}
Subsequently, the drift tubes are tested with respect to dark current. The specified limit is 2~nA at $10\%$ above the nominal operating voltage of 2730~V. Five drift tubes are tested simultaneously  
by first evacuating them to remove residual humidity, filling them with the nominal gas mixture at 3 bar absolute pressure, and applying the test voltage of 3015~V.
The dark current is measured continuously over time period of at least 10~minutes in which the dark current must stabilise below 2~nA without spikes within the first few minutes which indicate contamination or damage of the tubes or the sense wire. 

Drift tubes failing the criteria were flushed with argon gas for one minute. This successfully removed residual dust particles and humidity.
Drift tubes with stable dark current below 2~nA over 10~minutes were accepted for chamber assembly. No further conditioning of the drift tubes assembled in chambers was necessary after these procedures. The large majority of the drift tubes showed leakage currents below 0.5~nA, the sensitivity limit of the HV power supply. Only $1\%$ of the drift tubes failed the test and were discarded due to dark currents several orders of magnitude above the limit which was attributed to defects of the wire or the inner tube wall.

\subsection{Chamber Geometry Measurements}
\label{sec:QAQC:Geometry}
\subsubsection{Alignment Platform Positions}
\label{sec:QAQC:Geometry:Platforms}
The 3D platform positions and orientations relative to the sense wire grids at the chamber ends, have to be known  with a precision of better than $20~\mu$m and the platforms must be positioned within $\pm 200~\mu$m of their nominal positions to ensure the chamber alignment monitoring precision in the ATLAS muon spectrometer~\cite{ATLAS-TDR-10}. 
The magnetic field sensor platforms need to be positioned with the same precision within $500~\mu$m of their nominal position. 
The platform positions were verified and measured with high precision still on the chamber assembly jig. The platforms for different types of sensors platforms and their positions on the chambers are shown in Figure~\ref{fig:sMDT_chamber_CAD}. 

The platform positions were measured with an electro-mechanical feeler arm which measures positions in three dimensions over a range of more than a meter. The sensor positions and orientations on the platforms  are defined by reference surfaces and embedded steel spheres on the platforms which are measured by the feeler arm
with respect to the nearest assembly comb (see Figure~\ref{fig:gluingsetup}) in the chamber coordinate system. 
Each of the reference points is measured eight times to improve the statistical measurement precision while iteratively excluding outliers which are more than 50~$\mu$m from the mean. 
The resulting precision of the AP and B-field sensor platforms position measurements in the chamber coordinate system is better than 10~$\mu$m. For the CCC platforms, which are further away from the combs, the precision is around $20~\mu$m fulfilling the requirement. This translates into a platform angle measurement precision of  0.5~mrad. 

\begin{figure}[htb]
	\centering
	\subfloat[]{\includegraphics[width=0.8\textwidth]{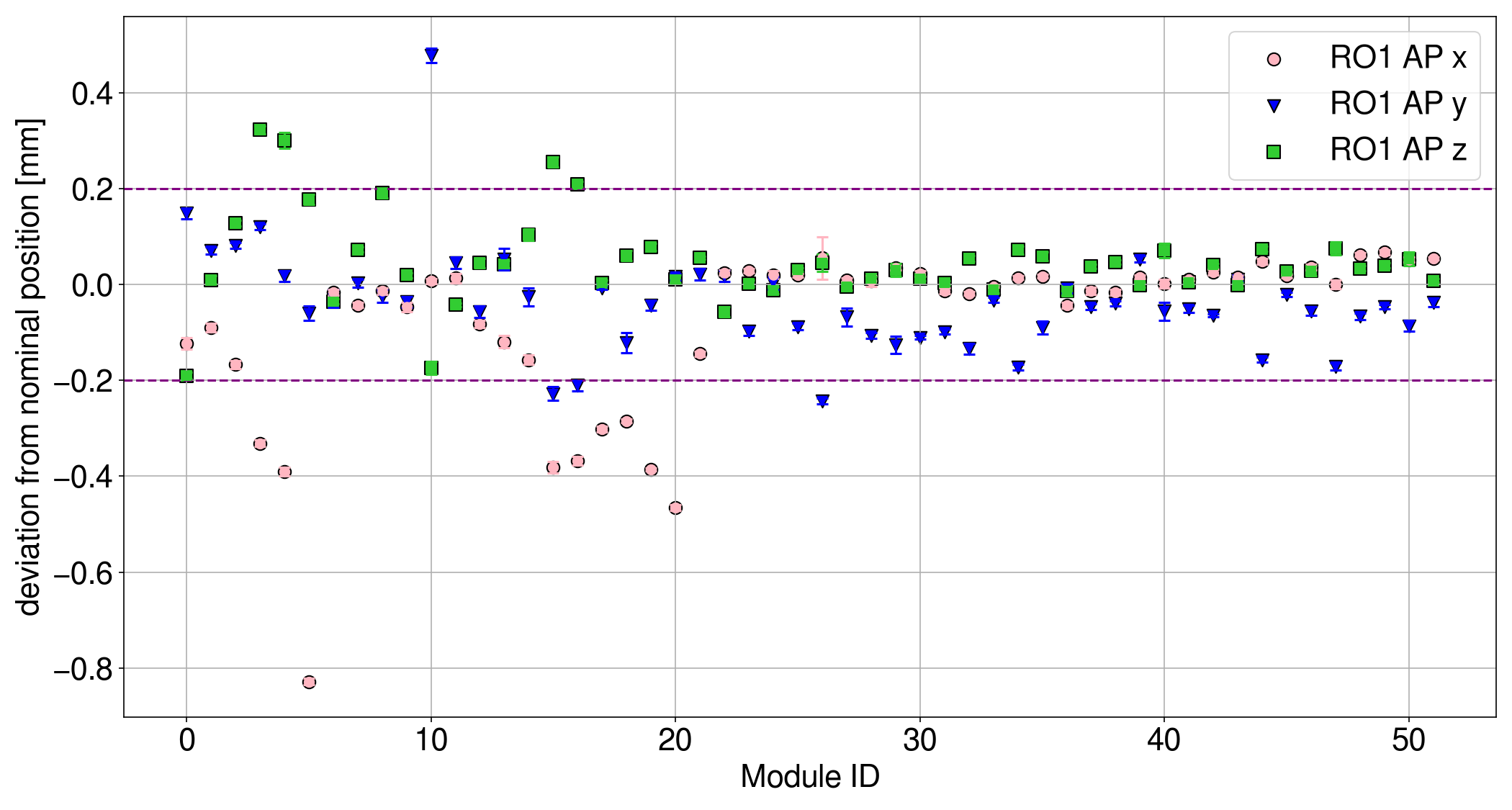}} \\
	\subfloat[]{\includegraphics[width=0.8\textwidth]{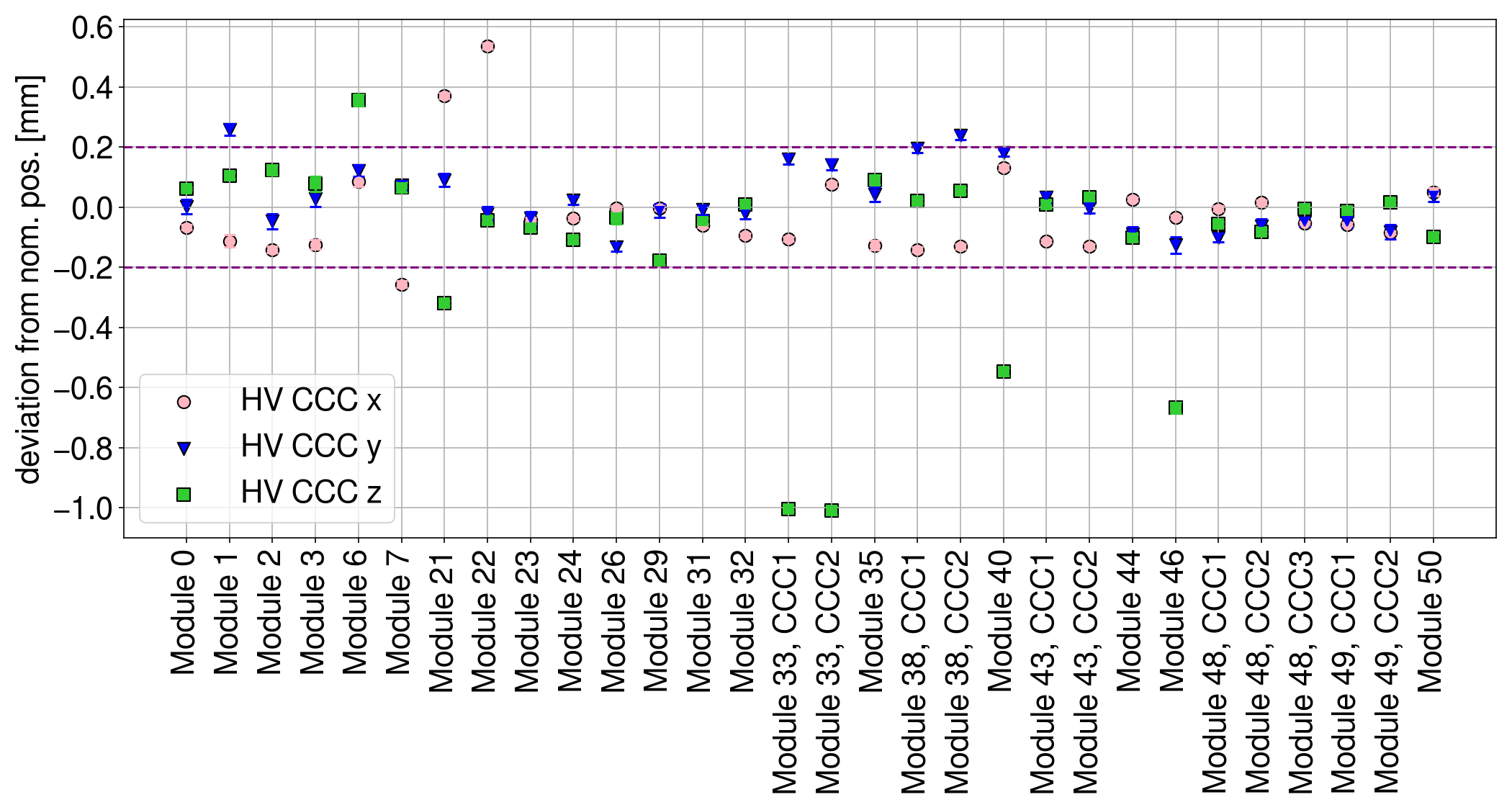}}
	\caption{Offsets of the platform positions of the chamber with respect to their nominal values for the produced BIS1-6 chambers in their production sequence for the examples of (a) the AP platforms in one chamber corner on the RO side and of (b) the CCC platforms on the HV side. The measurement precision is within the symbol sizes.}
	\label{Fig:AP_platforms_summary}
\end{figure}

Figure~\ref{Fig:AP_platforms_summary} shows examples of the offsets of the platform positions with respect to the nominal positions for the AP platforms in one corner of the chamber on the RO side and for the CCC platforms on the HV side depending on the chamber construction sequence. The large majority of the AP and CCC platforms are positioned within the required $\pm 200~\mu$m of their nominal position, and the magnetic field sensor platforms within $\pm 500~\mu$m. The measured platform angles are within a few mrad of the nominal values.
When occasionally increased displacements were observed, the jigging and mounting plates for platform positioning were readjusted. 

\subsubsection{Sense Wire Grid Measurements}
\label{sec:QAQC:Geometry:WirePos}
The assembled sMDT chambers, mounted in their transport frames on 3-point supports (two on the RO side and one on the HV side), are moved from the assembly table to the portal coordinate measuring machine (CMM) in the same clean room to measure the external reference surfaces of the drift tube endplugs and thus the sense wire positions separately at two the chamber ends. 

\begin{figure}[htp]
	\centering
	\includegraphics[width=0.6\textwidth]{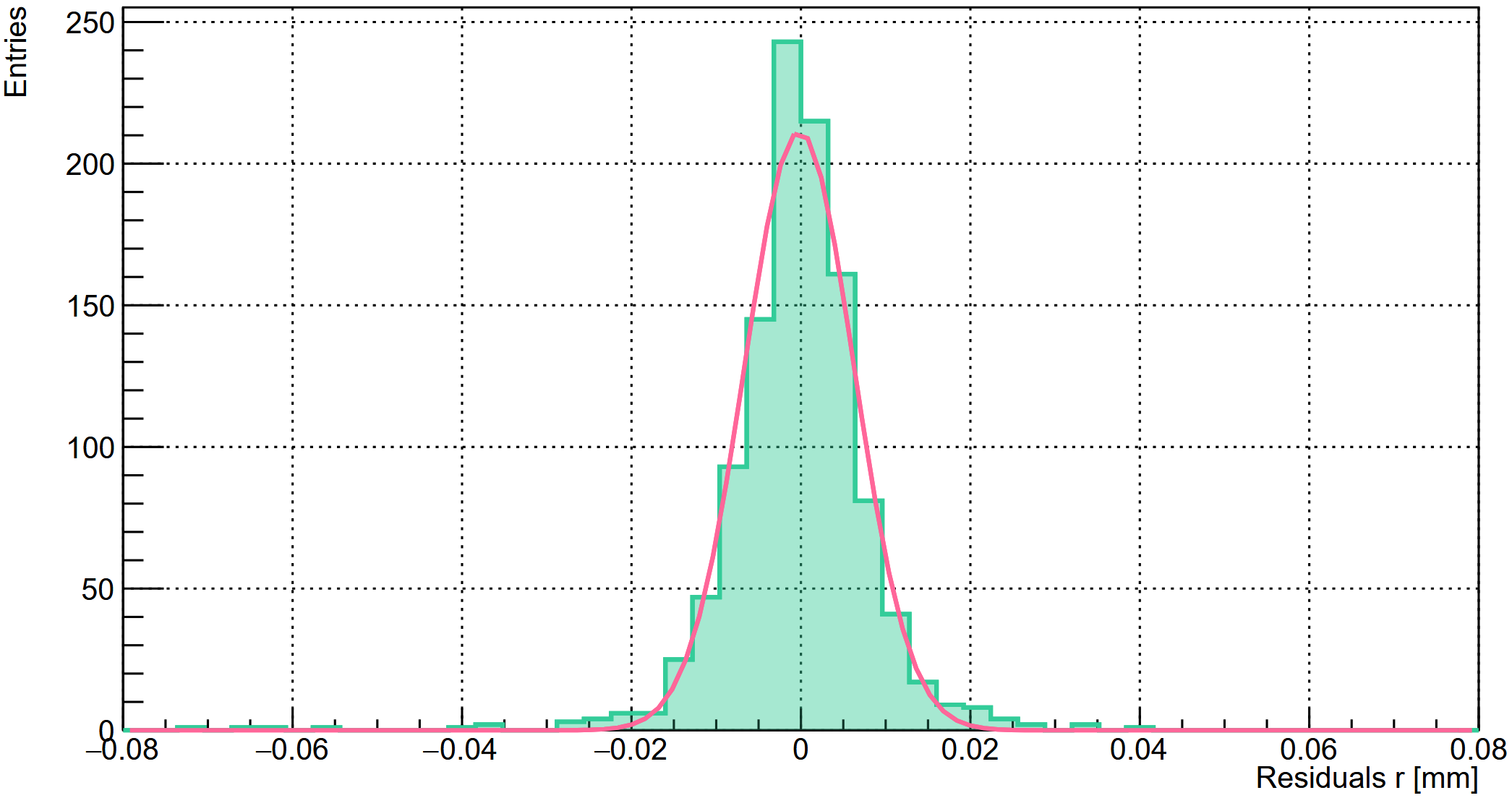}
	\caption{Distribution of the residuals in $r$ of the measured sense wire positions with respect to the combined site grid fit of RO and HV side of a BIS1-6 sMDT chamber.}
	\label{Fig:CMM_residuals}
\end{figure}

A regular grid is fitted to the measured sense wire positions on each chamber end and to both ends together by an automated application on the Web-interface, in order to determine the geometrical parameters of the chamber. The site grid parameters used for the muon track reconstruction in ATLAS are the wire pitches in $y$ and $z$ direction as well as the multilayer displacements in $y$ and $z$.  
The wire positioning accuracy is determined from the residuals of the measured wire positions with respect to the fitted grid in $y$, $z$, and the radial distance $r$ (see Figure~\ref{Fig:CMM_residuals}). The wire positions follow a Gaussian distributions to good approximation without systematic deviations. 

\begin{figure}[htp]
  \centering
  \includegraphics[width=0.8\textwidth]{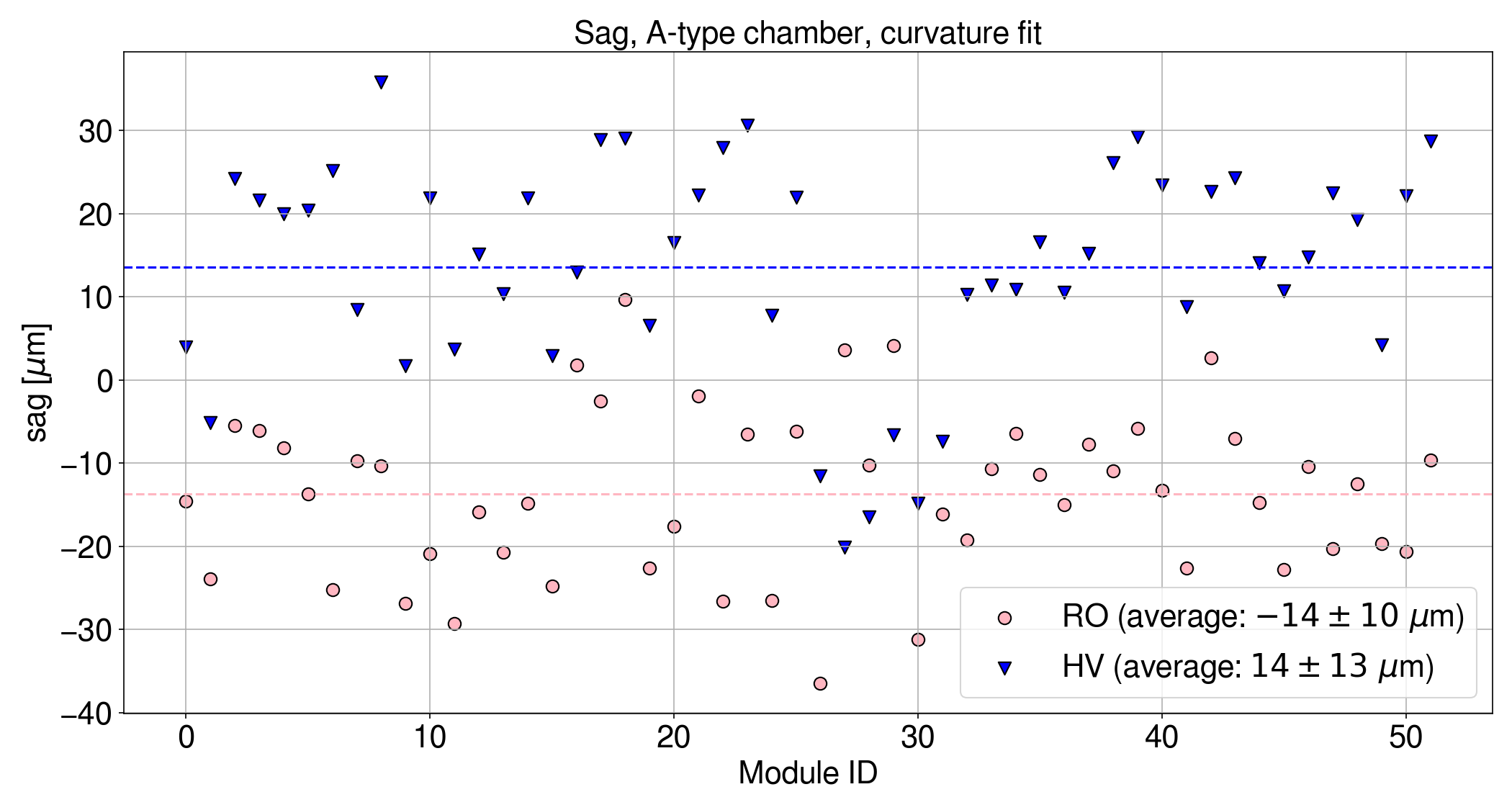}
  \caption{Fitted sagittas of gravitational deformations of the RO and HV sides of the produced chambers, appoximated by parabolas (curvature fit). The average values are indicated by dotted and dashed horizontal lines.}
  \label{fig:CMM_sags}
\end{figure}

In addition, the individual layer distances were fitted to detect possible construction mistakes before the assembly of the next chamber, as well as the gravitational deformations of the chamber ends, described by the sagittas of parabolas extending between the multilayer ends, to verify that they are negligible for the chamber precision. They are not corrected for in the quoted wire positioning accuracies. The fitted sags of the 
caused by the chamber being mounted on few points in the transport frame (see mounting points in Figure~\ref{fig:sMDT_chamber_CAD}). Figure~\ref{fig:CMM_sags} shows the fitted sagittas on both chamber ends over the whole chamber production. For most chambers, the sagittas have positive values on the HV side (parabolas opening downwards for a single chamber support in the middle) and negative values on the RO side (parabolas opening upwards for two supports at the ends) as expected.
\begin{table}[h]
    \caption{Construction site grid parameters averaged over the whole chamber production from fits to the measured wire grids on the RO and the HV side, compared to the measured comb parameters from Table~\ref{tab:CMM_combs}, as well as the wire positioning accuracies from the residual distributions}
    \centering
    \adjustbox{max width=\textwidth}{
    \begin{tabular}{|l|ccc|r|}
        \toprule
        Parameter               &  \multicolumn{3}{c|}{Average measured value}        &   Nominal value  \\
        \cline{2-4} 
                                &  RO+HV  & RO &  HV                         &                  \\
        \midrule
   		pitch $y$               & $13.090\pm0.002$~mm   & $13.089\pm0.003$~mm   & $13.091\pm0.003$~mm    &   13.076~mm      \\
   		pitch $z$               & $15.1001\pm0.0001$~mm & $15.1003\pm0.0001$~mm & $15.0998\pm0.0001$~mm  &   15.1000~mm     \\
   		multilayer distance $y$ & $45.590\pm0.004$~mm   & $45.593\pm0.006$~mm   & $45.588\pm0.006$~mm    &   45.600~mm      \\
   		multilayer distance $z$ & $1.9\pm4.8$~$\mu$m    & $3.6\pm4.5$~$\mu$m    & $1.3\pm6.4$~$\mu$m     &   0.000~$\mu$m     \\
        \hline
   		residual RMS $y$  & $11.7\pm2.2$~$\mu$m  & $10.6\pm1.4$~$\mu$m & $10.1\pm1.4$~$\mu$m   &       \\
   		residual RMS $z$  & $8.1\pm1.7$~$\mu$m   & $6.9\pm1.3$~$\mu$m  & $6.3\pm1.6$~$\mu$m    &       \\
   		residual RMS $r$  & $9.0\pm2.0$~$\mu$m   & $7.9\pm1.7$~$\mu$m  & $7.2\pm2.2$~$\mu$m    &       \\
   		residual $\sigma$ $y$   & $10.6\pm2.2$~$\mu$m  & $9.0\pm1.3$~$\mu$m& $9.0\pm1.3$~$\mu$m  &     \\
   		residual $\sigma$ $z$   & $7.0\pm1.5$~$\mu$m   & $5.8\pm1.2$~$\mu$m & $4.7\pm0.9$~$\mu$m   &   \\
   		residual $\sigma$ $r$   & $7.3\pm1.3$~$\mu$m   & $6.2\pm1.3$~$\mu$m & $5.2\pm0.9$~$\mu$m   &   \\
        \bottomrule
    \end{tabular}
    }\label{tab:CMM_parameters}
\end{table}
\begin{figure}[htp]
	\centering
	\subfloat[]{\includegraphics[width=0.8\textwidth]{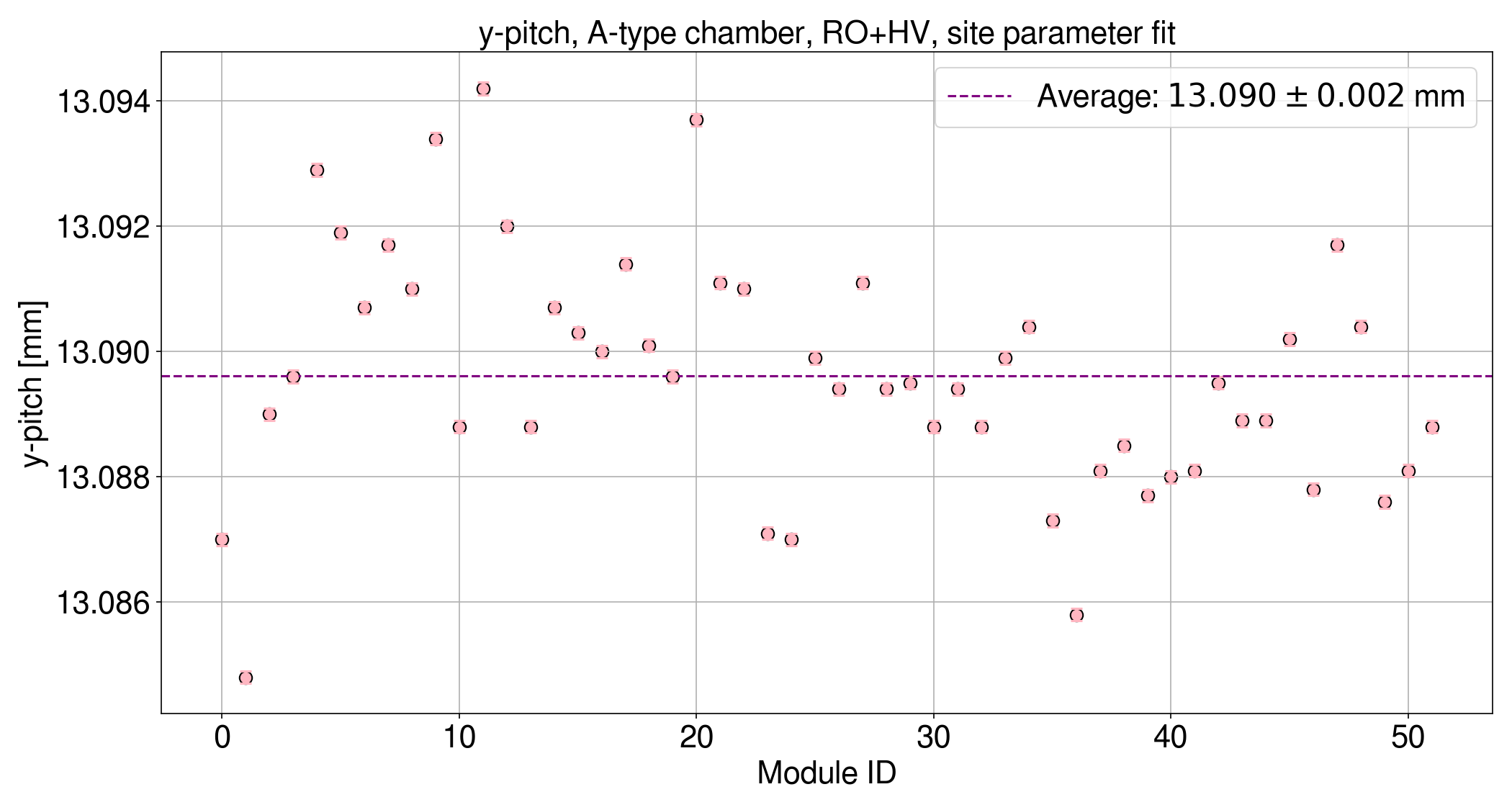}} \\
	\subfloat[]{\includegraphics[width=0.8\textwidth]{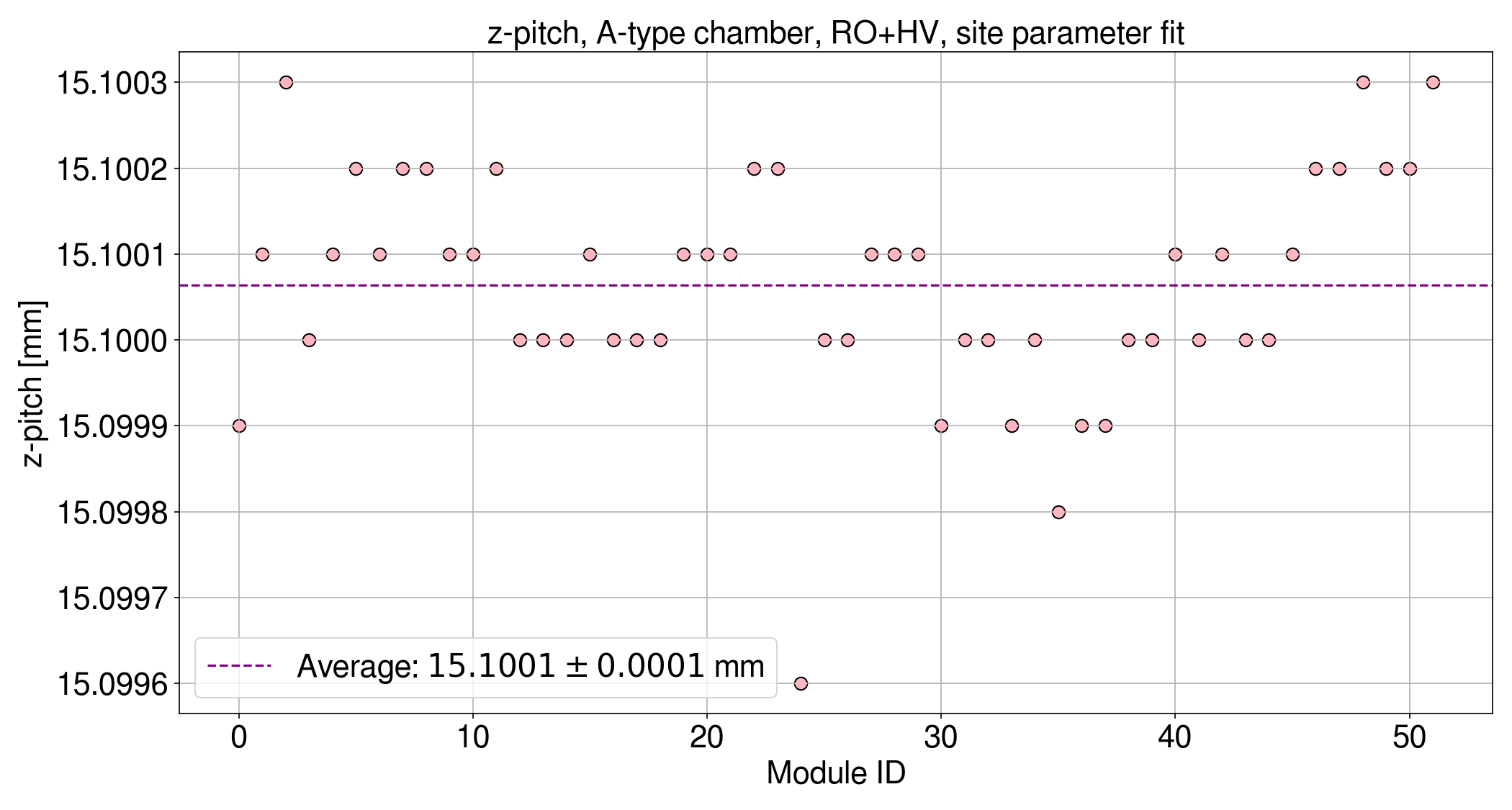}}
	\caption{Fitted wire grid parameters averaged over RO and HV side for all A-side chambers in the sequence of construction: a) $y$-pitch perpendicular to the chamber plane and (b) $z$-pitch within the chamber plane.
    The parameters are stable within tight margins over the whole production. 
    The averages over all chambers, the construction site parameters, are indicated by horizontal lines.} 
	\label{Fig:CMM_param_summary_plots_1}
\end{figure}
\begin{figure}[htp]
	\centering
	\subfloat[]{\includegraphics[width=0.8\textwidth]{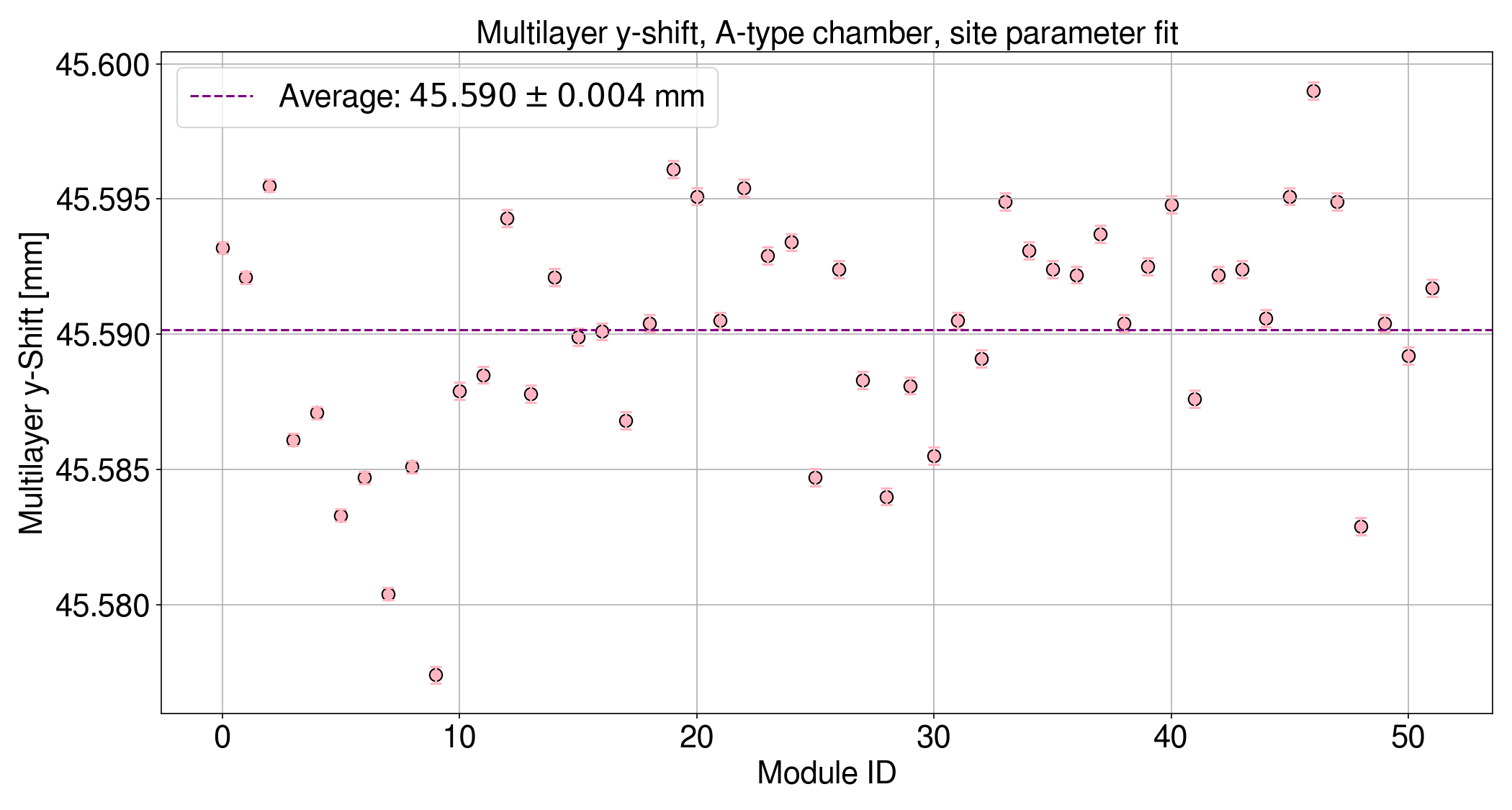}} \\
	\subfloat[]{\includegraphics[width=0.8\textwidth]{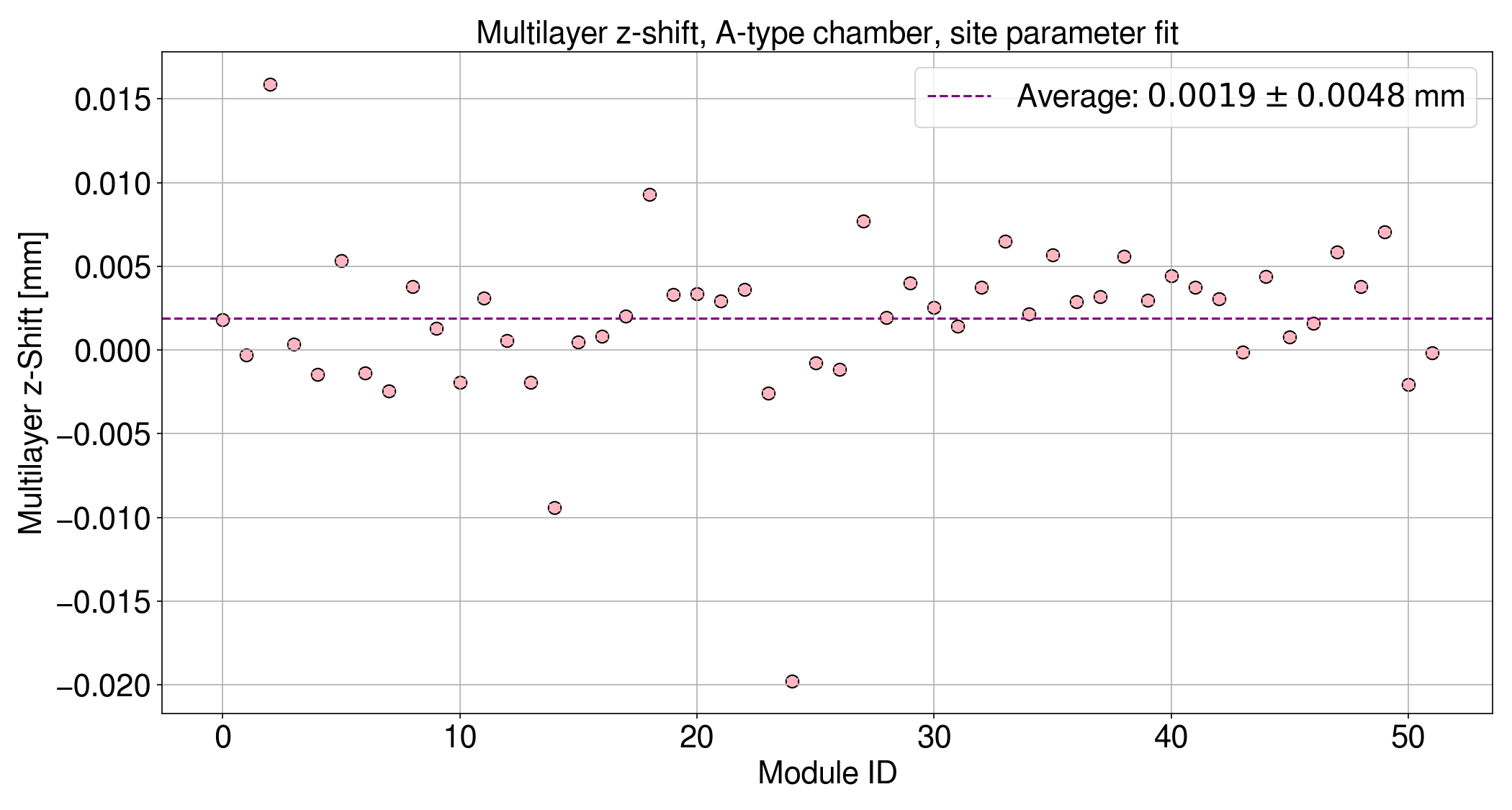}} 
	\caption{Fitted wire grid parameters averaged over RO and HV side for all A-side chambers in the sequence of construction: a) multilayer distance in $y$ direction perpendicular to the chamber plane and b) multilayer displacement in $z$ direction within the chamber plane. The parameters are stable within tight margins over the whole production. The averages for all chambers, the construction site parameters, are indicated by horizontal lines.}
	\label{Fig:CMM_param_summary_plots_2}
\end{figure}
Table~\ref{tab:CMM_parameters} shows the production averages of the fitted construction site grid parameters for RO and HV side combined and of the RMS values and standard deviations of the corresponding residual distributions in $y$, $z$ and $r$. The average wire positioning accuracy is $7.3~\mu$m, significantly below the required precision of 20~$\mu$m. Figures~\ref{Fig:CMM_param_summary_plots_1} and~\ref{Fig:CMM_param_summary_plots_2} show the fitted wire array parameters for all sMDT chambers during the production at MPI.

\begin{figure}[htp]
	\centering
	\includegraphics[width=0.8\textwidth]{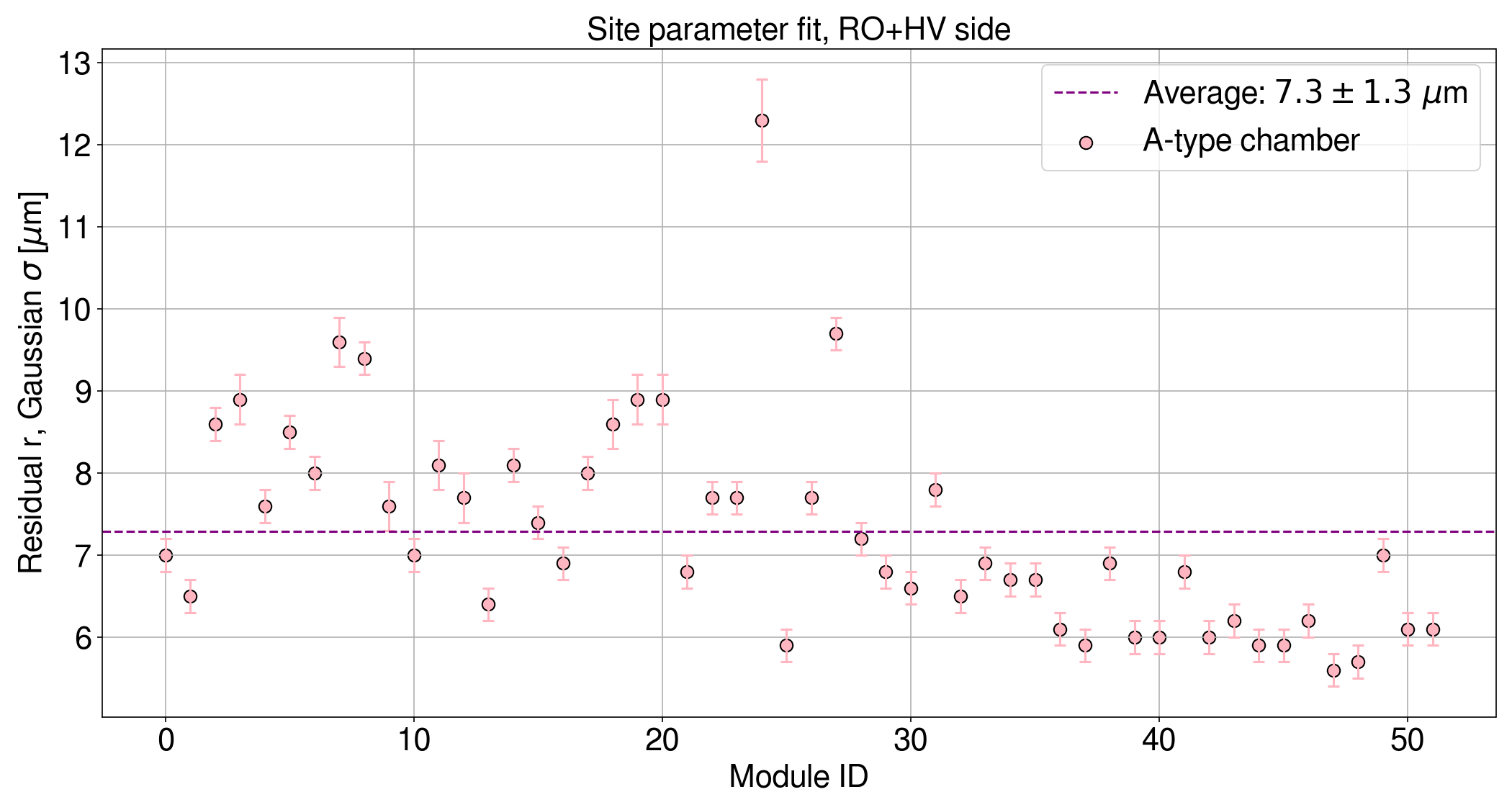}
	\caption{Wire positioning accuracy as standard deviation of $r$ residual distribution for the combined site parameter fit of the RO and HV sides. Production averages are shown as dashed lines. One can see that the chamber precision still improved over the course of the production.}
	\label{Fig:CMM_res_summary_plots}
\end{figure}

The wire positioning accuracies for all A-type sMDT chambers are shown in Figure~\ref{Fig:CMM_res_summary_plots}. The wire positioning accuracies are consistently under 10~$\mu$m except for a single chamber during the production. Later in the production, wire positioning accuracies of around 6~$\mu$m were the norm. \par

\subsubsection{Chamber Torsion}
\label{sec:QAQC:Geometry:torsion}
The CMM was used to validate the measurement the in-plane alignment system of the chamber torsion between RO and HV side around the tube axis. Due to their small spacer height, the BIS1-6 sMDT chambers are particularly susceptable to such torsion deformations which depend on the external forces, are varying with their locations and orientations in the ATLAS detector and thus have to be monitored. For the torsion measurement, both the RO and the HV side wire positions were measured simultaneously in the same coordinate system.  
\begin{figure}[htb]
  \centering
  \includegraphics[width=0.8\textwidth]{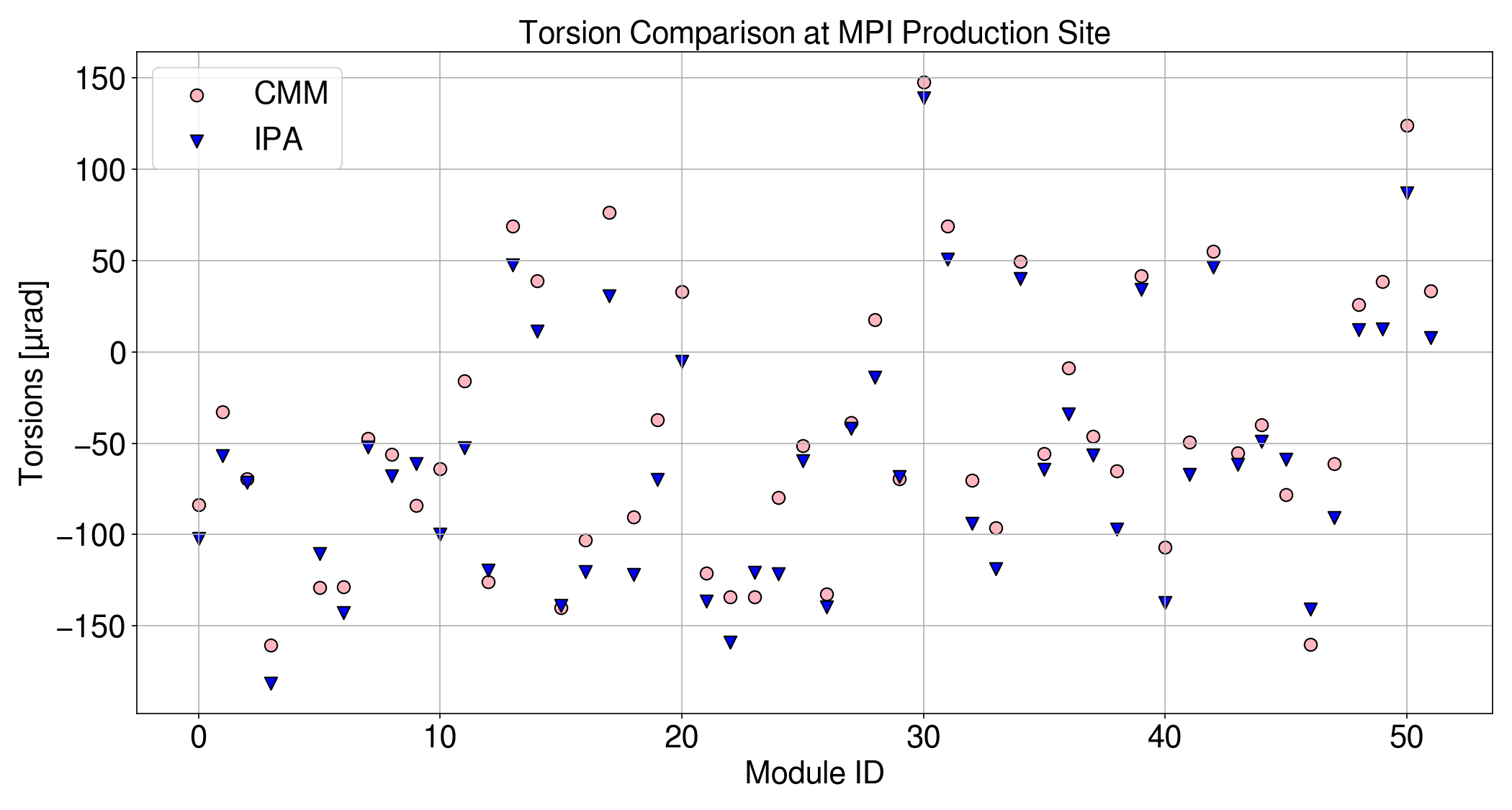}
  \caption{Relative torsion angles between RO and HV side for all produced chambers as measured by the in-plane alignment system (IPA) and by the CMM.}
  \label{Fig:Torsion}
\end{figure}

\subsection{Chamber Gas Leak Rate Measurement}
\label{sec:QAQC:Gas}
The gas tightness of the assembled chambers is validated by measuring the pressure drop in each multilayer with the nominal gas mixture $\text{Ar}:\text{CO}_2$ (93:7) from 3 bar absolute over at least 24 hours in a temperature-controlled room with a precision of 0.2 mbar while the gas temperature is monitored with a precision of 0.2~K with the temperature sensors distributed over the two multilayers.
The gas leakage rate is required to be less than 
$2 N_{\text{tubes}}\cdot 10^{-8}~\text{bar}\cdot\text{liter}/\text{s}$ at a reference temperature of 20°C, where 
$2 N_{\text{tubes}}$ is the number of endplugs in the multilayer. This requirement corresponds to a pressure drop per hour of $0.28~\text{mbar}/\text{h}$. At this stage, before the installation of the readout electronics and Faraday cages, there is still access for repairs of the gas distribution system. 

\begin{figure}[htb]
  \centering
  \includegraphics[width=0.8\textwidth]{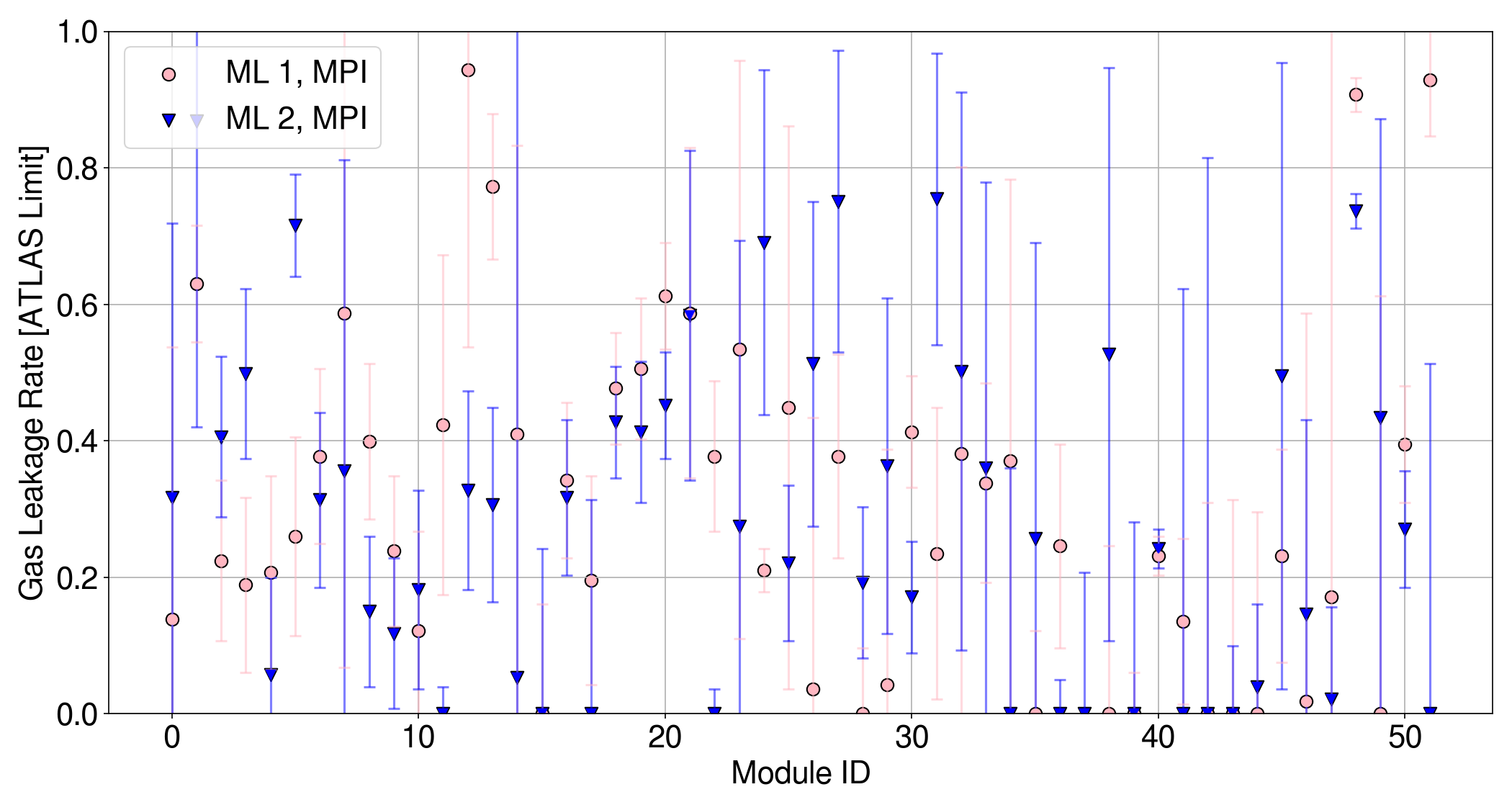}
  \caption{Gas leak rates as a fraction of the required upper limit 
  of the two multilayers of the produced A-side BIS1-6 chambers measured during chamber production.}
  \label{fig:leakrate_summary}
\end{figure}

\begin{figure}[htb]
  \centering
  \includegraphics[width=0.8\textwidth]{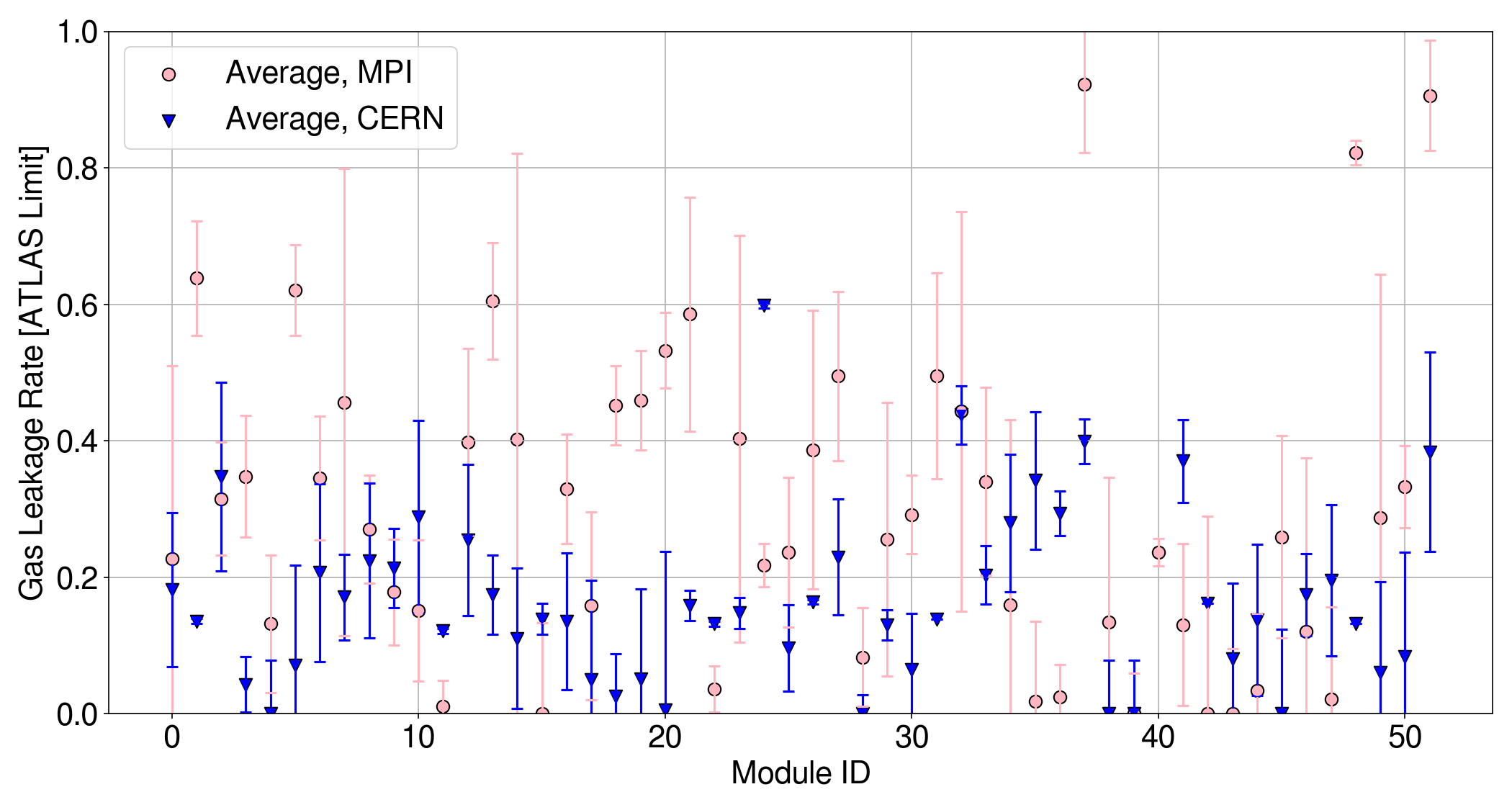}
  \caption{Gas leak rates as a fraction of the upper limit of the produced A-side BIS1-6 chambers (average of the two multilayers) measured at the construction site and after the shipment to CERN. At CERN the certification at the production site was confirmed with higher precision (see text) showing that the leak rate not only did not increase due to the transport, but is actually smaller than estimated before, in general about a factor of 5 below the limit.}
  \label{fig:leakrate_BB5_summary}
\end{figure}

After the first certification, the gas leak rate is measured again after the installation of the electronics cards on the endplugs during cosmic ray data taking (see next chapter) and after the transport of the chamber to CERN. Figure~\ref{fig:leakrate_summary} shows the gas leak rates as a fraction of the required upper limit for the produced chambers measured at the production site. All chambers fulfilled the requirement. The precision is governed by the time over which pressure drop has been measured. During chamber production this time span was limited to 24 hours in order to maintain the production rate of two weeks per chamber. The pressure drop measurement over several months between chamber completion and shipment to CERN allowed for confirmation of the previous measurements with much higher precision (see Figure~\ref{fig:leakrate_BB5_summary}).

\subsection{Chamber Performance with Cosmic Muons}
\label{sec:QAQC:Cosmics}
After completion of chamber assembly including readout electronics and Faraday cages, the performance of the chambers including dark currents and electronics noise rates has been tested in a cosmic ray teststand with a scintillation trigger counter covering the whole chamber width. A set of prototype mezzanine cards (see Section~\ref{sec:Chamber:Electronics}), sufficient for one chamber, with new ASD chips~\cite{ABOVYAN2019374,ASD2_IEEESensor,ASD2_manual} from the pre-production for HL-LHC (see Section~\ref{sec:Chamber:Electronics}) was used for the performance tests.
As verification of the full functionality of the chambers, the muon detection efficiencies and spatial resolutions of the drift tubes have been measured and their stability monitored over the production period. 

\begin{figure}[htb]
  \centering
  \includegraphics[width=0.8\textwidth]{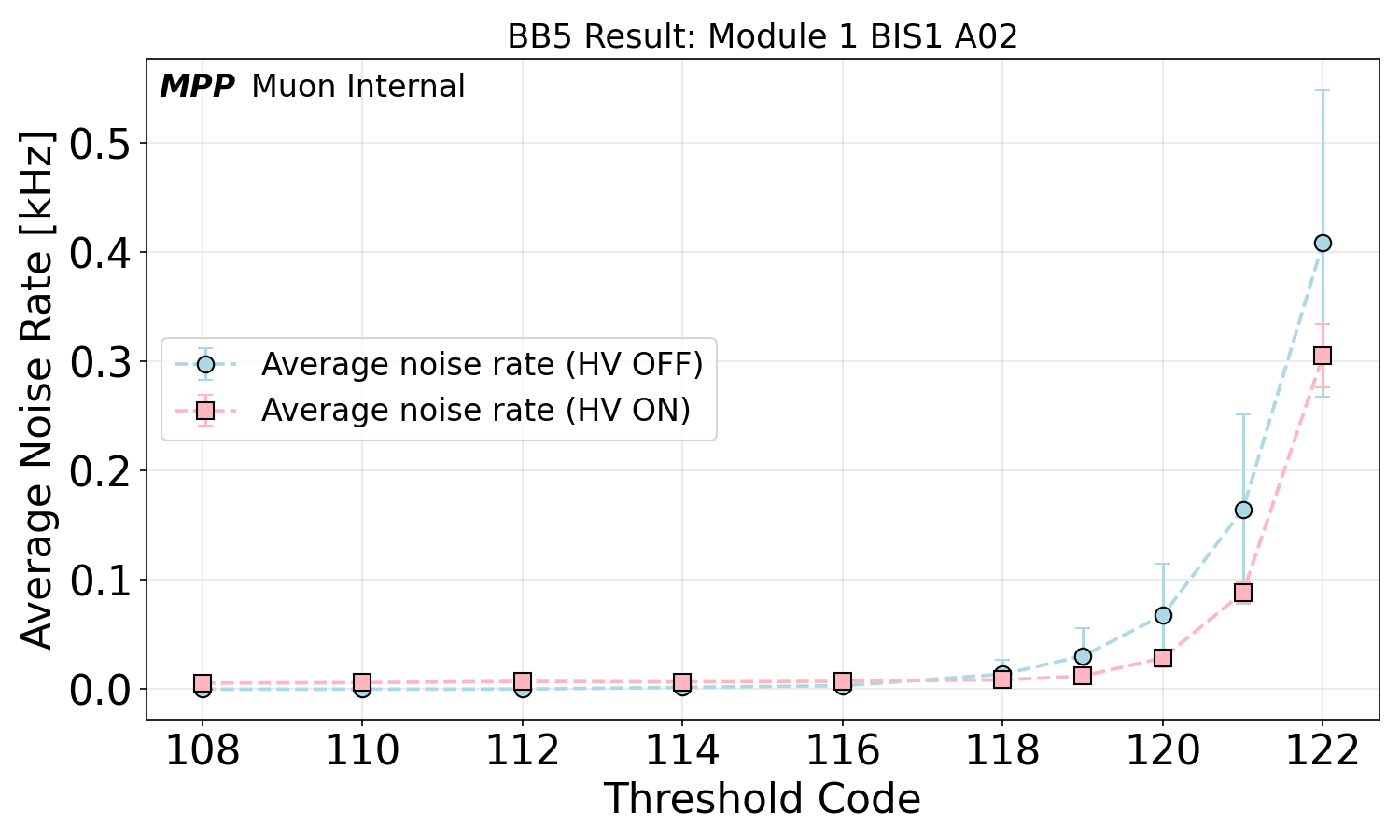}  
  \caption{Threshold scan for the Phase~2 ASD chip with and without HV turned on. The average electronics noise rate of the channels of a chamber is measured as a function of the discriminator threshold setting for hysteresis code 14 (see text). The effective threshold decreases with increasing threshold code. The working point with threshold code 114 was chosen well below noise turn-on for safe operation in ATLAS and thus also for chamber commissioning. With HV turned on, cosmic muon hits are registered in addition to the electronics noise.}
  \label{fig:threshold_scan1}
\end{figure}

The 8-bit discriminator threshold code of the ASD chips~\cite{ASD2_manual} (see Section~\ref{sec:Chamber:Electronics}), ranging from 0 to 255 with physical thresholds increasing with  decreasing code numbers from the zero at 127.5, and the 4-bit hysteresis code ranging from 0 to 14 were set to 114 and 14, respectively, used in previous testbeam studies~\cite{testbeam_1} and corresponding to an effective threshold of 22 primary ionization electrons according to the calibration in~\cite{testbeam_2} for the ASD chips from pre-production used for the chamber testing at the construction sites. The settings have been optimised in a measurement of the noise rate as function of the threshold code for fixed hysteresis code 14 (see Figure~~\ref{fig:threshold_scan1}).

\subsubsection{Noise Rate Measurements}
\label{subsec:noiseRates}
\begin{figure}[htb]
  \centering
  \includegraphics[width=0.8\textwidth]{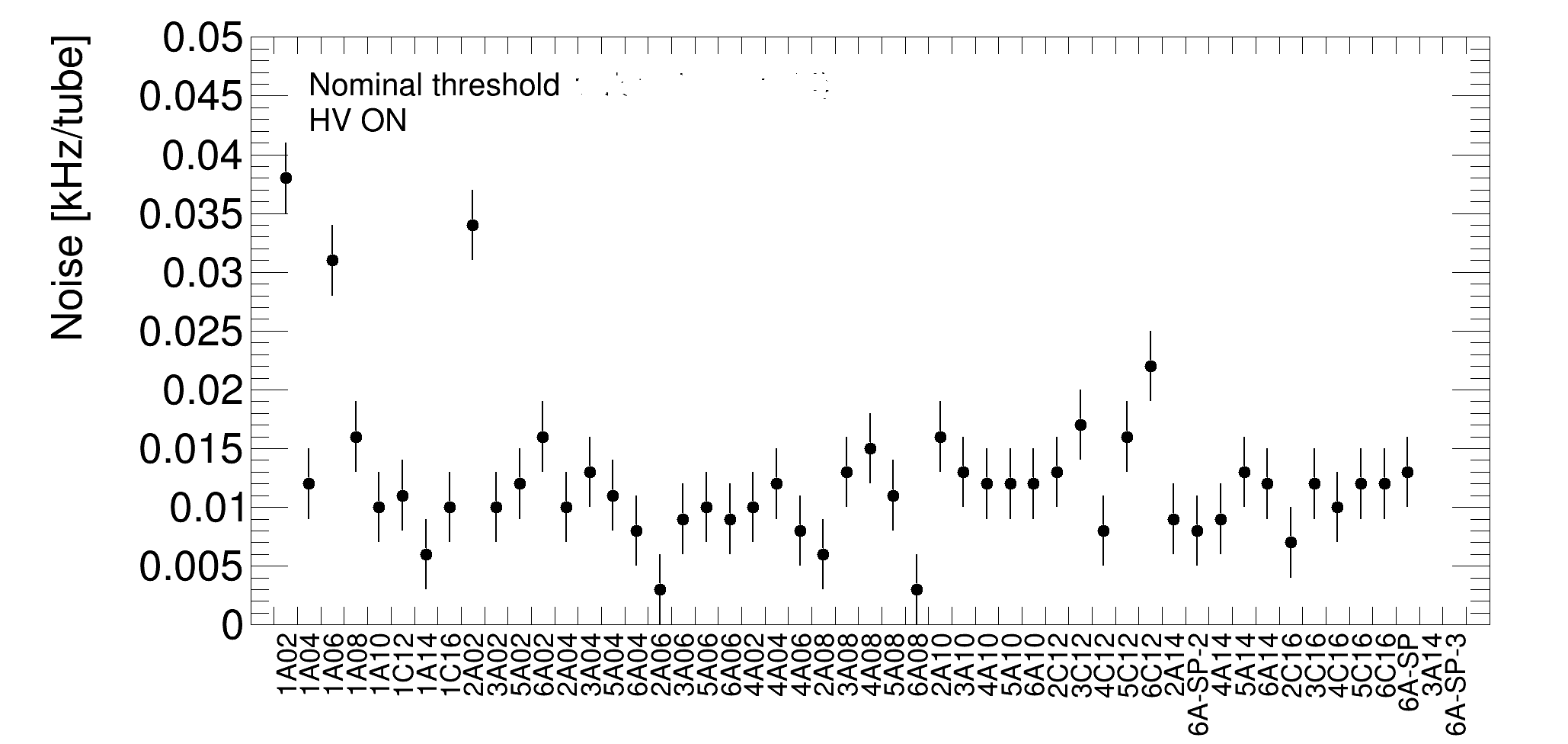}
  \caption{Average drift tube noise rates per chamber for all A-side BIS1-6 sMDT chambers in the sequence of production for HV turned on at the nominal threshold and hysteresis settings corresponding to an effective threshold of 22 primary electrons.}
  \label{fig:noise_rates_mpi}
\end{figure}

The noise rates are measured using an 800~Hz random trigger for HV turned on  and off. In the first case, cosmic muon events are also registered at a rate of about 10~Hz per tube.
The noise rate of each tube is determined as
\begin{equation*}
f_{\text{noise}}=\frac{N_{\text{events}}}{\Delta t_{\text{RO}}\cdot N_{\text{triggers}}},
\end{equation*}
with $N_{\text{events}}$ the number of registered events, $N_{\text{triggers}}$ the number of triggers during the measurement, and 
$\Delta t_{\text{RO}}$ the readout time window set to $1.3~\mu$s.
The requirements for the certification of a chamber were less than 1~kHz noise rate per channel and less than 100~Hz average noise rate per channel which leaves plenty of margin in case of additional noise sources during operation in the ATLAS detector. 
The noise rates were first measured at a lower effective threshold of 15 primary electrons corresponding to threshold and hysteresis settings of 120 and 14, respectively, to identify channels potentially susceptable to noise using the same criteria and perform improvements to drift tube and electronics grounding in few cases. Afterwards the chambers were certified to fulfill the standard noise rate requirements.
Figure~\ref{fig:noise_rates_mpi} shows the average drift tube noise rates  per chamber with HV turned on at the nominal threshold settings for the A-side sMDT chambers.  
The noise rates are a factor of 5 to 10 below the acceptance limit of 100~Hz per channel, at the level of the expected cosmic counting rate, uniformly over the whole production time.

\subsubsection{Drift Tube Efficiency and Spatial Resolution}
\label{subsec:tubeEff}
\begin{figure}[htb]
  \centering
  \includegraphics[width=0.8\textwidth]{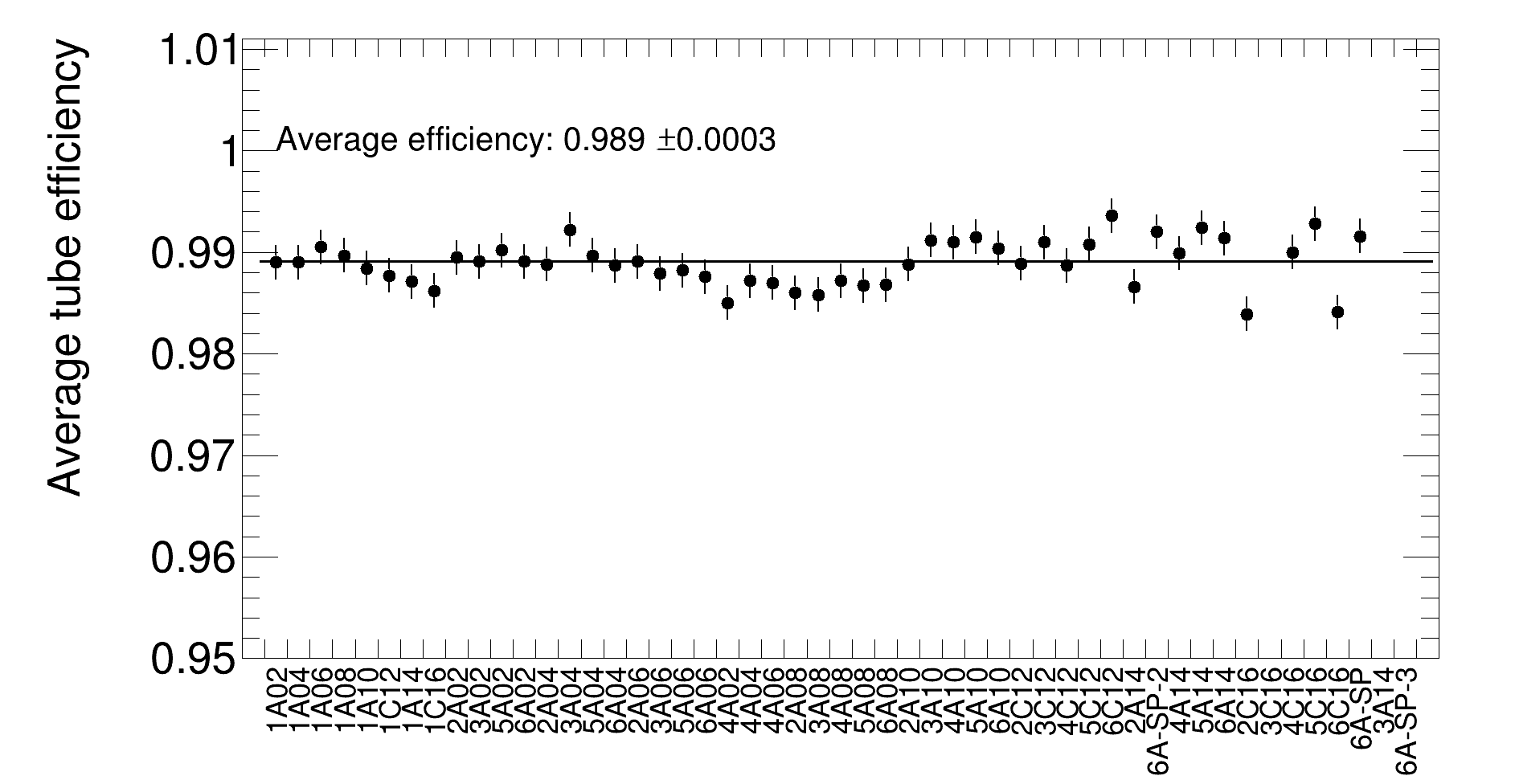}
  \caption{Average drift tube muon hit efficiencies per chamber inside the active gas volume for all A-side BIS1-6 sMDT chambers in the sequence of production at the nominal threshold settings.}
  \label{fig:efficiency}
\end{figure}

\begin{figure}[htb]
  \centering
  \includegraphics[width=0.8\textwidth]{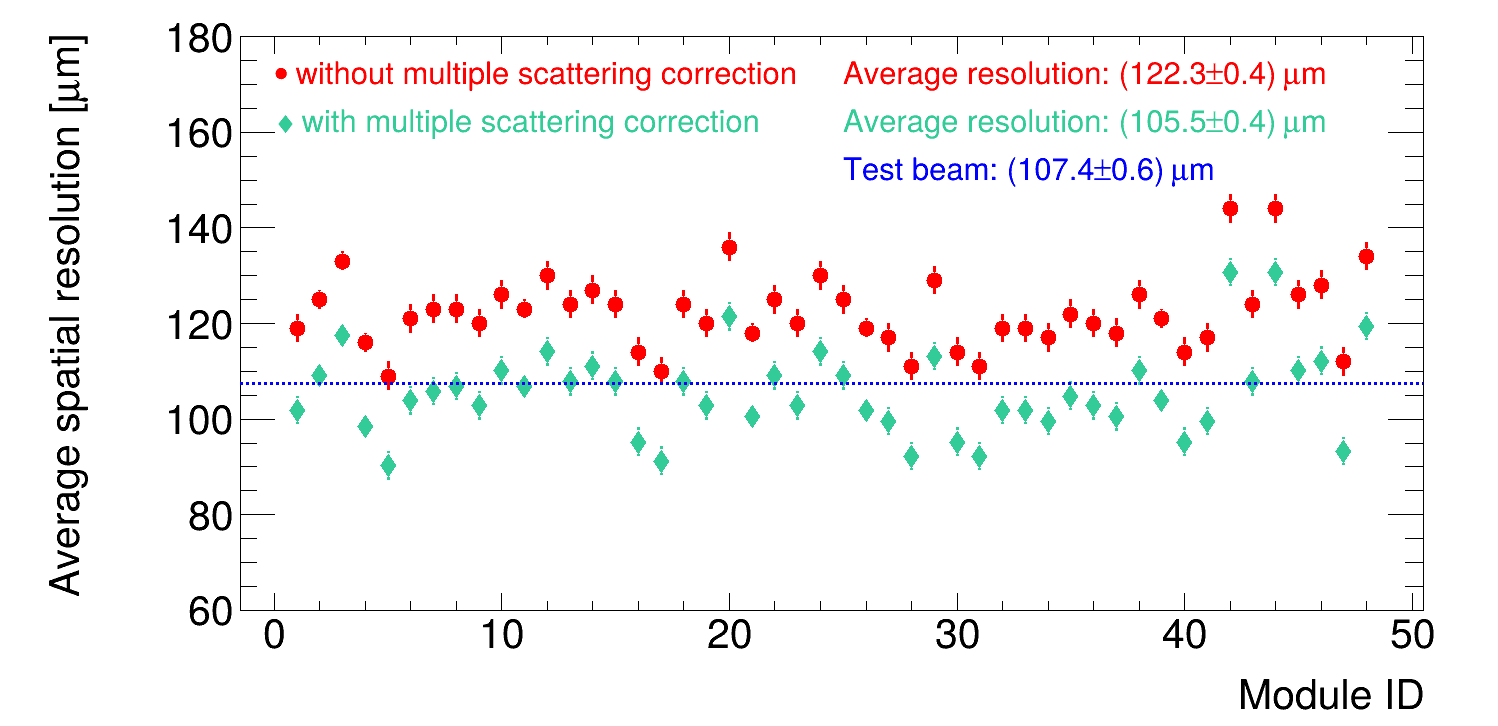}
  \caption{Average drift tube spatial resolution per chamber for the BIS1-6 A-side chambers in the sequence of production for the nominal threshold settings with effective threshold (from calibration in~\cite{testbeam_2}) of 22 primary electrons with (green diamonds) and without (red circles) multiple scattering corrections, and without time walk corrections (see text). The results have been obtained with cosmic ray muons at the production site as part of the quality control test program before shipment of the chambers to CERN. The error bars do not include uncertainties in the multiple scattering corrections. The latter are confirmed by comparison with results from sMDT testbeam measurements (blue dotted line) in a high-energy muon beam at CERN using the same ASD readout chips developed for operation at HL-LHC and the same threshold settings. The results are in agreement with resolution measurements for all chambers at CERN with the final readout boards for HL-LHC~\cite{BIS_commissioning_CERN}.}
  \label{fig:resolution_BB5}
\end{figure}

The cosmic ray data collected over at least 24~h at the nominal threshold settings were used to determine the efficiency and spatial resolution of the chambers. Muon tracks were reconstructed from the recorded hits requiring at least 4~hits per track. A detailed description of the analysis is given in~\cite{Hadzic}. The drift tubes show uniform muon hit efficiency over 
the whole serial production (see Figure~\ref{fig:efficiency}) with an average of $0.989\pm0.0003$ inside the active volume. In total only 22 non-functional tubes, mostly due to broken sense wires, have been found and disconnected from gas supply, HV and readout electronics, two of them after shipment to CERN. They were not taken into account in the efficiency calculation.
The resolution was evaluated applying the method described in~\cite{Hadzic}. 
The ASD chips were used in the time-over-threshold mode. Thus no time-walk corrections were applied. The effect of the
multiple Coulomb scattering effect for the energy spectrum and angular distribution of the cosmic muons within the trigger acceptance was estimated by Monte Carlo simulation of the chamber and the test stand to amount to about $60~\mu$ which was quadratically subtracted in the resolution results shown in Figure~\ref{fig:resolution_BB5}. The obtained average drift tube spatial resolutions are uniform over the whole chamber production and in agreement between the measurements at the production site and after shipment of the chambers to CERN in 2023. The multiple scattering correction introduces the largest global systematic uncertainty not taken into account in the error bars.
Measurements in an energetic muon beam with negligible multiple scattering effects and with the same ASD chips and threshold settings are therefore also shown as reference for the resolution achievable with the new ASD chip for HL-LHC, $95\pm 1.0~\mu$m for the nominal threshold settings with time slewing corrections and without background irradiation~\cite{sMDT_electronics_performance}.

\section{Conclusions}
\label{sec:Conclusion}
The design and the construction of the new small-diameter Muon Drift Tube (sMDT) precision tracking chambers for the upgrade of the innermost barrel layer of the ATLAS Muon Spectrometer at HL-LHC at MPI Munich (BIS1-6 A-side chambers) have been discussed, including the procedures and results of extensive quality control tests of the individual drift tubes and of the assembled chambers. 
The 48 chambers plus 4 spares constructed in the years 2020 to 2022 all fulfill the requirements for installation in ATLAS. The average gas leak rate is about a factor of 5 below the specified limit. A construction rate of two weeks per chamber has been maintained over the whole chamber production.
The noise rate with the new front-end electronics developed for HL-LHC is exceptionally low, at the level of the cosmic muon counting rate, at the nominal threshold settings corresponding to an effective threshold of 22 primary electrons. The muon hit efficiency and spatial resolution have been measured for all drift tubes in the chambers and are uniform over the whole chamber production.

The achieved sense wire positioning accuracy and the resulting wire grid parameters as well as the positions of the optical alignment sensor platforms with respect to the sense wire grid have been measured using an automated coordinate measuring machine  
and an electro-mechanical feeler arm, respectively. These mechanical measurements are possible due to the particular design of the sMDT drift tubes which also improves the chamber assembly speed and precision compared to the legacy ATLAS MDT chambers. An average sense wire positioning accuracy 
of $7~\mu$m has been achieved compared to $20~\mu$m required and achieved for the MDT chambers.




\bibliographystyle{JHEP}
\bibliography{biblio.bib}


\end{document}